\documentclass[acmsmall]{acmart}

\usepackage{booktabs} 
\usepackage{enumerate}
\usepackage{graphicx}
\usepackage{amsmath}
\usepackage{multirow}
\usepackage{footnote}
\usepackage{threeparttable}
\usepackage{textcomp}
\usepackage{balance}
\usepackage{lipsum}
\usepackage{hyperref}
\usepackage{nicefrac}
\usepackage{xcolor}

\usepackage[ruled,vlined,linesnumbered]{algorithm2e}
\usepackage{amsmath}

\newcommand{\figref}[1]{Fig.~\ref{#1}}

\newcommand{\secref}[1]{Section~\ref{#1}}

\newcommand{\mycolor}{black}
\newcommand\newtext[1]{\textcolor{black}{#1}}

\usepackage{natbib}

\usepackage{caption}
\usepackage{subcaption}
\usepackage{listings}

\definecolor{codegreen}{rgb}{0,0.6,0}
\definecolor{codegray}{rgb}{0.5,0.5,0.5}
\definecolor{codepurple}{rgb}{0.58,0,0.82}
\definecolor{backcolour}{rgb}{0.95,0.95,0.92}

\lstdefinestyle{mystyle}{
  backgroundcolor=\color{backcolour}, 
  commentstyle=\color{codegreen},
  keywordstyle=\color{magenta},
  numberstyle=\tiny\color{codegray},
  stringstyle=\color{codepurple},
  basicstyle=\ttfamily\footnotesize, 
  breakatwhitespace=false,    
  language=Python,
  breaklines=true,                 
  captionpos=b,                    
  keepspaces=true,                 
  numbers=left,                    
  numbersep=5pt,                  
  showspaces=false,                
  showstringspaces=false,
  showtabs=false,                  
  tabsize=2,
  float=tp
}
\usepackage[colorinlistoftodos]{todonotes}

\usepackage{tikz}
\newcommand*\circled[1]{\tikz[baseline=(char.base)]{
            \node[fill=black,text=white,shape=circle,draw,inner sep=0.5pt] (char) {\normalsize{#1}};}}
\usepackage{colortbl}

\usepackage{pifont}

\definecolor{lightgray}{gray}{0.9}

\usepackage{algpseudocode}

\AtBeginDocument{%
  \providecommand\BibTeX{{%
    \normalfont B\kern-0.5em{\scshape i\kern-0.25em b}\kern-0.8em\TeX}}}

\setcopyright{acmcopyright}
\copyrightyear{2025}
\acmYear{2025}

\acmJournal{JACM}
\acmArticle{1}
\acmMonth{3}

\begin{document}

\title{MetaML-Pro: Cross-Stage Design Flow Automation for Efficient Deep Learning Acceleration}



\author{Zhiqiang Que}
\email{z.que@imperial.ac.uk}
\affiliation{%
  \institution{Imperial College London}
  \country{UK}
}




\author{Jose G. F. Coutinho}
\email{gabriel.figueiredo@imperial.ac.uk}
\affiliation{
  \institution{Imperial College London}
  \country{UK}
}

\author{Ce Guo}
\affiliation{
  \institution{Imperial College London}
  \country{UK}
}

\author{Hongxiang Fan}
\affiliation{
  \institution{Imperial College London}
  \country{UK}
}

\author{Wayne Luk}
\email{w.luk@imperial.ac.uk}
\affiliation{%
  \institution{Imperial College London}
  \country{UK}
}

\renewcommand{\shortauthors}{Zhiqiang Que, et al.}

\begin{abstract}

This paper presents a unified framework for codifying and automating optimization strategies to efficiently deploy deep neural networks (DNNs) on resource-constrained hardware, such as FPGAs, while maintaining high performance, accuracy, and resource efficiency. Deploying DNNs on such platforms involves addressing the significant challenge of balancing performance, resource usage (e.g., DSPs and LUTs), and inference accuracy, which often requires extensive manual effort and domain expertise. Our novel approach addresses two core issues: (i)~encoding custom optimization strategies and (ii)~enabling cross-stage optimization search. In particular, our proposed framework seamlessly integrates programmatic DNN optimization techniques with high-level synthesis (HLS)-based metaprogramming, leveraging advanced design space exploration (DSE) strategies like Bayesian optimization to automate both top-down and bottom-up design flows. Hence, we reduce the need for manual intervention and domain expertise. In addition, the framework introduces customizable optimization, transformation, and control blocks to enhance DNN accelerator performance and resource efficiency. \textcolor{\mycolor}{We further formalize a cross-stage, constrained Bayesian optimization procedure that couples predicate–action bottom-up feedback (via BRANCH) with FORK/REDUCE order search, enabling automated selection, ordering, and tuning across software and HLS tasks.}
Experimental results demonstrate up to a 92\% DSP and 89\% LUT usage reduction for select networks, while preserving accuracy, along with a 15.6-fold reduction in optimization time compared to grid search. These results highlight the potential for automating the generation of resource-efficient DNN accelerator designs with minimal effort, 
\textcolor{\mycolor}{ resulting in large resource savings with bounded exploration cost.} 
\end{abstract}



\maketitle

\section{Introduction}
\label{sec:intro}

The field of deep learning has witnessed unprecedented growth in recent years, driven by the increasing demand for efficient and high-performance applications~\cite{lecun2015deep, que2021accelerating, wojcicki2022accelerating, lou2025low}. Consequently, FPGA-based deep neural network (DNN) accelerator design and optimization have gained significant attention~\cite{que2022reconfigurable, que2020mapping, que2020optimizing, que2022remarn}. The development of an efficient FPGA-based DNN design requires a diverse skill set that combines expertise in machine learning with low-level knowledge of the target hardware architecture~\cite{sze2017efficient, que2020reconfigurable}. Optimizing these DNN designs is a complex process, as it involves balancing competing objectives~\cite{que2025trustworthy}. On one hand, high accuracy during inference is crucial from an application perspective. On the other hand, the design must be optimized for the underlying hardware architecture, meeting power, latency, and throughput requirements while fitting into FPGA devices~\cite{que2024ll,  jedi-linear}. Achieving an optimal balance between these conflicting objectives requires careful consideration and effective optimization strategies~\cite{que2025trustworthy}. 

Existing optimization techniques for DNNs and hardware have a limitation in that they are often handled in separate stages, with information exchanged only in a top-down manner from the application to hardware. Bottom-up information flow, where hardware design insights inform DNN model optimization, typically requires manual intervention. However, this information exchange across multiple stages can significantly enhance the optimization process, as application and hardware optimizations can interact and affect each other. For instance, modifying the DNN model architecture could impact the FPGA resource utilization, and optimizing the FPGA design could influence the DNN model's accuracy~\cite{wang2019deep, que2024ll}. In addition, there has been limited focus on combining optimization strategies targeting different abstraction levels, such as algorithmic level neural networks and High Level Synthesis (HLS) C++, hindering the selection of the most effective combination of optimization techniques for a given problem. In summary, only a limited number of studies have thoroughly explored the vast configuration space that emerges when combining DNN optimization techniques across both software and hardware domains.

In this paper, we address these aforementioned technical challenges as follows:

\begin{itemize}
\item \textbf{C1. Custom Co-Optimization Strategies.} First, this work employs a unique co-optimization approach that refines both the DNN software model and the hardware architecture. By encoding software and hardware optimizations independently and composing them as needed, this approach enables modular reuse and adaptation of optimizations across different benchmarks and platforms. Through programmatic manipulation, it ensures tight coordination between DNN model and hardware-level optimizations, significantly expanding the design space and enhancing exploration capabilities.

 \item \textbf{C2. Cross-Stage Optimization Search.} Second, developing efficient DNN accelerators requires a coherent strategy that automates the selection, combination, and tuning of optimization techniques across multiple abstraction levels~\cite{ney2021half, abdelfattah2020best, yang2019synetgy,zhang2022algorithm}. It is challenging to identify the most effective combination, order and tuning of optimization techniques to ensure optimal outcomes~\cite{zhang2020dnnexplorer, xu2020autodnnchip}. This work addresses these challenges by employing a cross-stage optimization search approach that coordinates the interaction between different stages of the design flow. By leveraging a combination of top-down and bottom-up optimization strategies, the framework dynamically adapts to the specific requirements of the hardware and the application, guiding the optimization process iteratively and identifying the best optimization techniques and their sequences. 
 \end{itemize}

\begin{figure}
\begin{center}
\includegraphics[width=1.00\linewidth]{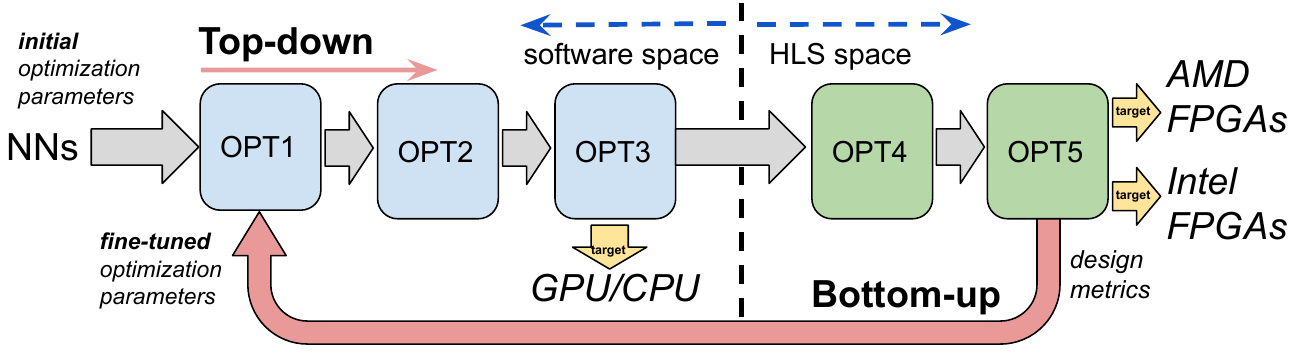}
\end{center}
   \caption{The proposed approach.}
\label{fig:approach}
\end{figure}

Our proposed co-optimization framework, as illustrated in Fig.~\ref{fig:approach}, operates across various computational spaces, such as software and High-Level Synthesis (HLS), and can target \textcolor{\mycolor}{diverse FPGA platforms} (see \textbf{C1}). Each computational space represents an autonomous, custom optimisation task (e.g., OPT-1 to OPT-5, and potentially more), governed by tunable parameters. Moreover,
this work utilizes a Bayesian~\cite{mehrabi2020bayesian, tuli2023codebench, que2025trustworthy} guided nested-loop optimization structure involving both local and global phases, combining very distinct optimization tasks and incorporating feedback from the latter stages of the design flow to refine the process iteratively (see \textbf{C2}). \textcolor{\mycolor}{Finally, while we instantiate Bayesian optimization for the outer DSE, the controller is algorithm-agnostic and can host evolutionary or RL-based search as drop-in alternatives.}

We make the following contributions in this paper:
\begin{itemize}
\item A novel co-optimization framework that enables modular and adaptive hardware-software exploration for FPGA-based DNN accelerators. It independently encodes and combines optimizations for efficient design refinement. 

\item \textcolor{\mycolor}{A cross-stage optimization algorithm. We formalize a nested-loop, constraint-aware Bayesian optimization procedure over software- and HLS-level optimization tasks. The algorithm uses predicate–action bottom-up feedback (via BRANCH) to adapt accuracy/utilization tolerances on constraint violations, and FORK/REDUCE to search task permutations and aggregate by Pareto dominance. This automates selection, ordering, and tuning across abstraction levels. }

\item A library of reusable optimization, transformation, and control tasks designed to be customizable and flexible, and that can be easily integrated into our co-optimization framework. This library includes optimization tasks such as pruning, quantization, and scaling to enhance DNN accelerator performance and resource efficiency. Moreover, we leverage metaprogramming to programmatically analyze and modify HLS C++ designs, enhancing design flexibility, performance tuning, and resource efficiency. Some of the tasks in our library are specific to certain applications and target technologies, while others remain agnostic, providing versatility and adaptability within the framework.

\item A comprehensive evaluation of the proposed framework using multiple benchmarks and different optimization strategies. This evaluation provides insights into the effectiveness of the framework and its optimization modules under different scenarios (Section~\ref{sec:evaluation}).
\end{itemize}

To the best of our knowledge, this is the first automated framework that combines reusable optimization components with adaptive feedback and global exploration to jointly balance resource efficiency and accuracy in deep learning designs.

\subsubsection*{Relationship to Prior Publications}

This paper expands on our previous studies~\cite{que2023metaml, que2024deep, que2024optimizing} which primarily focus on local optimizations. In~\cite{que2023metaml}, we propose MetaML, an optimization framework, that customizes design flow for DNNs with local optimizations. In~\cite{que2024deep}, we introduce control capabilities within the optimization framework, while \cite{que2023metaml} focuses on Quantization Heuristic Search (QHS) optimization task for compressing DNN accelerators. 
However, these prior studies are limited in their ability to perform global optimization due to their focus on isolated optimization tasks. The inability to account for interdependencies between optimization tasks across stages mean that the overall design might not have achieved the optimal balance between accuracy, performance, and hardware resource. This work addresses these limitations by incorporating bottom-up feedback and leveraging Bayesian optimization to enable global optimization across the entire design flow. By adjusting optimization tasks at each stage based on performance feedback, our framework ensures that each optimization choice contributes to a more cohesive and globally optimal solution. This extended approach maintains high accuracy while improving resource efficiency, specifically in resource-constrained environments.

\section{Related Work}\label{sec:related_work}

The field of FPGA-based DNN acceleration has rapidly expanded, leading to the development of numerous optimization techniques and tools. 
Several co-optimization techniques have been proposed that optimize both algorithm and hardware stages for DNNs on FPGAs, as discussed in various papers~\cite{yang2019synetgy, zhang2022algorithm, hao2018deep, hao2019fpga, hao2019nais,  hao2020effective, jiang2020hardware, dong2021hao, hao2021enabling} and in hardware-aware neural architecture search studies like~\cite{ney2021half, abdelfattah2020best, fan2022algorithm}. 
While hardware-aware neural architecture search (HW-NAS) approaches typically involve a holistic approach where the search algorithm must consider both software and hardware aspects together, often requiring a broader range of expertise within both NAS and hardware, our work modularizes the optimization process, allowing domain experts to focus on their areas of expertise. 
In addition, this modular design allows for individual optimization tasks to be reused independently in different contexts or different models, while the holistic nature and tight coupling of HW-NAS makes it challenging to reuse parts of the optimization process independently.
This work can independently update and refine specific modules without affecting the entire system, resulting in better reusability and adaptability.

Other approaches offer end-to-end software frameworks, such as Xilinx's Vitis AI~\cite{kathail2020xilinx} and Intel's OpenVINO~\cite{OpenVINO2023},  which optimize DNNs with pre-built optimizations for deployment on specific target technologies. However, they are not designed for easy addition of new optimization strategies or to search for the most effective combined strategies.

Furthermore, some frameworks allow developers to describe and customize DNN optimization strategies, but their scope is limited. For instance, ScaleHLS~\cite{ye2022scalehls} and SOTA-OPT~\cite{agostini2022mlir} focus on HLS-based optimization strategies and hardware designs, but their optimization scope is restricted to MLIR. TVM~\cite{chen2018tvm} is a general-purpose DNN compilation framework that offers performance portability across different types of devices, but optimizations occur solely at the graph (IR) level. \texttt{da4ml}~\cite{sun2025da4ml} is an open-source library that compiles machine-learning models into hardware designs. It targets fully pipelined implementations with an initiation interval (II) of 1, applies distributed arithmetic (DA) optimizations, and can emit synthesizable Verilog and VHDL for multiple FPGA back-ends.

Moreover, frameworks, such as FINN~\cite{umuroglu2017finn}, HLS4ML~\cite{schulte2025hls4ml}, and fpgaConvNet~\cite{venieris2016fpgaconvnet}, provide optimized hardware building blocks for FPGA-based DNN accelerators, and allow optimization strategies to be codified using these blocks. However, they do not support automated bottom-up optimization flows, where hardware stage information guides the optimization of the application (DNN) stage.

To overcome the limitations of existing optimization techniques for DNNs and hardware, our framework provides fully automated high-level optimization flows backed by reusable target-specific and target-agnostic building blocks. More specifically, our approach enables automated top-down and bottom-up flows, allowing for the creation of customized cross-stage optimization strategies for various DNN designs. Moreover, our approach integrates new and existing optimization techniques targeting various levels of abstraction, such as neural networks through graph optimizations and HLS C++ through source-to-source optimizations~\cite{fccm20_artisan}.
\section{Our Approach} \label{sec:approach}

\subsection{Overview}

In this paper, we introduce a novel framework for creating customizable design-flows that optimize FPGA-based DNNs, with a focus on automating both top-down and bottom-up flows between application and hardware optimization stages.

\begin{figure}
\begin{center}
\includegraphics[width=1.00\linewidth]{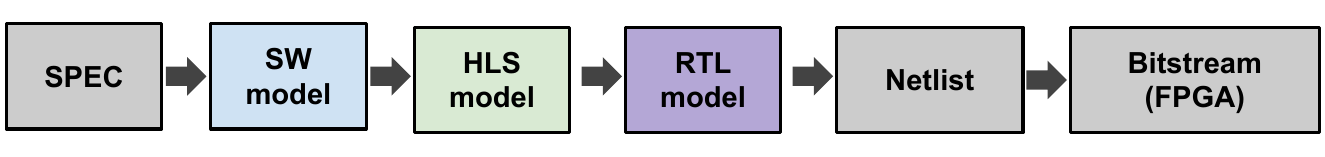}
\end{center}
   \caption{A typical FPGA design flow. This FPGA design flow begins by defining the specifications (SPEC), followed by implementing and testing the design in software (SW). If High-Level Synthesis (HLS) is used, high-level code (e.g., C/C++) is converted into hardware description language (HDL). The design then moves to the Register Transfer Level (RTL) stage, where it is described in HDL (e.g., Verilog/VHDL) and synthesized into a netlist of logic gates. Finally, the netlist is used to generate a bitstream, which is loaded onto the FPGA to configure the hardware. Each stage progressively refines the design from concept to implementation.}
\label{fig:designflow}
\end{figure}

A design-flow represents a sequence of tasks that convert a high-level specification into a final hardware design. A design-flow is typically implemented as a multi-stage pipeline, with each stage operating on a specific model abstraction. As the pipeline progresses, the model abstraction is gradually reduced and optimized, taking into account the features of the target device. For example, a typical design-flow targeting FPGAs is shown in Fig.~\ref{fig:designflow}. It begins with defining specifications (SPEC) and testing the design in software (SW). If High-Level Synthesis (HLS) is used, a high-level DNN model, such as one described in TensorFlow, can be translated into a C++ HLS model using tools like HLS4ML. This model is then converted into an RTL description (e.g., Verilog/VHDL) using tools like Vivado HLS. The RTL model is further synthesized into a netlist, which is optimized for the FPGA’s architecture, and finally translated into a bitstream that configures the FPGA hardware.

Feedback information can be utilized to enable lower-stage information to refine higher-stage optimizations, employing more accurate metrics that were previously unavailable. We have developed a framework that allows users to codify fully automated optimizing design-flows with the following requirements:

\begin{itemize}

\item \textbf{Customizable}: Refers to the ability for users to customize, extend, or modify the design-flow to meet specific needs and support experimentation. Users can select a set of building tasks that implement optimization, transformation, or control tasks, combine them in a specific order, and fine-tune their parameters to generate a specific optimization flow. Additionally, they can create their own tasks and integrate them into their design-flow (see~\textbf{C1});

\item \textbf{Multi-level}: Refers to the scope of optimizations and the ability to target multiple levels of abstraction, typically associated with different stages of a design-flow. For instance, optimizations operating at the Neural Network level are performed at the graph-level, while optimizations at the HLS C++ level are performed using source-level transformations (see~\textbf{C1});

\item \textbf{Cross-stage}: Refers to the capability of a design-flow to support top-down and bottom-up flows between optimization stages, such as software and hardware, respectively (see~\textbf{C2}).

\end{itemize}

This work focuses on custom optimizations in the SW and HLS stages while using default configuration for the remaining stages, however, the same concepts can be extended to the other stages, which is left for future work.

\subsection{Architecture}

\begin{figure}[tp]
   \centering
    \includegraphics[width=0.8\linewidth]{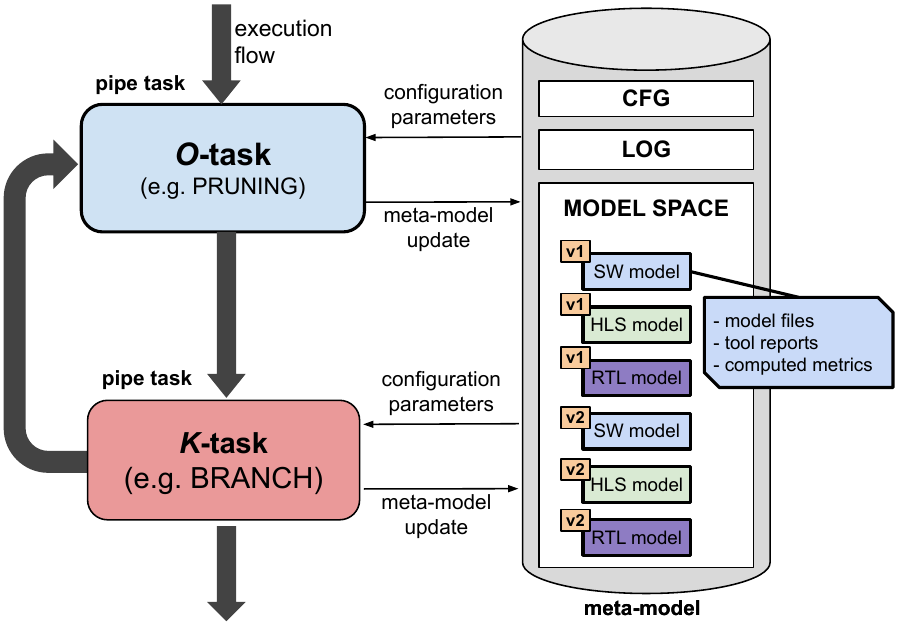}
  \caption{A connection between a $O$-task and a $K$-task. A pipe task has a uniform interface allowing any two pipe tasks to be connected (although there may be constraints about how many connections a task can handle). A $O$-task typically enhances DNN models based on specific objectives and constraints. A $K$-task on the other hand, manages the control flow. Each connection defines a unidirectional stream between a source task and a target task.}
  \label{fig:pipetasks} 
\end{figure}

 Fig.~\ref{fig:pipetasks} illustrates the key architectural components of a programmatically generated design flow using our approach, which comprises two main components: the pipe task and the meta-model.

The pipe task is the fundamental building block of the design flow and is responsible for implementing specific a task, such as optimization or control. By connecting pipe tasks, we  create a complete design flow. The design flow's internal architecture is represented by a cyclic directed graph, where nodes represent tasks, and edges represent \textbf{paths} between tasks. A path represents a dependency between two tasks, such that a task can only be executed if its dependencies have finished executing.

The meta-model, on the other hand, is responsible for storing all of the design flow's states, including the configuration parameters of each pipe task and their respective outputs. Instead of communicating directly with each other, pipe tasks share information through the meta-model, which serves as a shared space. The meta-model has three sections, as illustrated in Fig.~\ref{fig:pipetasks}(b): configuration, log and model space.

The configuration section (\textbf{CFG}) is a key-value store that keeps the parameters of all pipe tasks in the design flow. These parameters can configure all pipe tasks of the same type or specific instances, and they can be supplied by the user or automatically modified by the pipe tasks as part of an automated tuning process.
The log section (\textbf{LOG}) stores the runtime execution trace of the design flow. This section can be useful for debugging and understanding the execution flow of the design.

Finally, the \textbf{model space} stores the models generated during the execution of a design flow. Each stored model is versioned, and models derived by different stages can coexist in the same model space. In the example shown in the figure, we have stored six models covering three abstraction levels: DNN, HLS C++ , and RTL. Each model in the model space encapsulates all its supporting files, tool reports, and computed metrics. Moreover, they can be accessed and manipulated by any pipe task in the design flow, enabling cross-stage optimizations.

 \begin{table}[pt]
\centering
\caption{ A list of implemented pipe tasks.}
\vspace{5pt}
\label{table:pipetasks}
\scalebox{1.0}{
\begin{tabular}{l|c|c|l}
\hline
\rowcolor[HTML]{C0C0C0} 
\textbf{Type} &
  \textbf{Role} &
  \textbf{Multiplicity} &
  \textbf{Parameters}\\ \hline
JOIN            & $K$  & many-to-1 & - \\ \midrule
BRANCH          & $K$  & 1-to-2    & fn: meta-model $\rightarrow$ bool  \\ \midrule
FORK            & $K$  & 1-to-many &  - \\ \midrule

REDUCE
& $K$  
& many-to-1 
& \begin{tabular}[c]{@{}l@{}} fn: [meta-model] \\ $\rightarrow$  meta-model \end{tabular} 
\\ \midrule

STOP &
  $K$ &
  1-to-0 &
  fn: meta-model $\rightarrow$  output
  \textbf{} \\ \midrule

HLS4ML  
  & $\lambda$ 
  & 1-to-1 & 
       \begin{tabular}[c]{@{}l@{}}
        default\_precision \\ 
        IOType \\
        FPGA\_part\_number \\
        clock\_period \\
        test\_dataset 
        \end{tabular}\\ \midrule
        
VIVADO-HLS 
  & $\lambda$ 
  & 1-to-1 & 
       project\_dir \\ \midrule
        
INTEL-oneAPI 
  & $\lambda$ 
  & 1-to-1 & 
       project\_dir \\ \midrule
       
\begin{tabular}[c]{@{}l@{}}
KERAS-MODEL\\-GEN 
\end{tabular}
  & $\lambda$ 
  & 0-to-1 & 
       \begin{tabular}[c]{@{}l@{}}
        train\_en \\
        train\_test\_dataset\\
        train\_epochs
        \end{tabular}
  \\ \midrule

PRUNING &
  $O$ &
  1-to-1 &
\begin{tabular}[c]{@{}l@{}}
        tolerate\_acc\_loss ($\alpha_p$) \\ 
        pruning\_rate\_thresh ($\beta_p$)\\
        train\_test\_dataset\\
        train\_epochs
        \end{tabular} 
        \\ \midrule
SCALING         
  & $O$ 
  & 1-to-1 & 
       \begin{tabular}[c]{@{}l@{}}
        default\_scale\_factor \\ 
        tolerate\_acc\_loss ($\alpha_s$)\\ 
        scale\_auto \\
        max\_trials\_num \\ 
        train\_test\_dataset\\
        train\_epochs
        \end{tabular} \\ \midrule

QUANTIZATION         
  & $O$ 
  & 1-to-1 & 
      \begin{tabular}[c]{@{}l@{}}
        tolerate\_acc\_loss ($\alpha_q$)\\ 
        train\_test\_dataset
        \end{tabular} \\ \bottomrule

\end{tabular}
}
\end{table}

\subsection{Reusable Pipe Tasks}

Table~\ref{table:pipetasks} presents a list of pipe tasks that have been implemented in our approach, along with their roles, multiplicity, and parameters. The multiplicity denotes the number of input and output channels that a task can handle. We classify pipe tasks into three types based on their roles:

\begin{itemize}

\item \textbf{$K$-task}: These are generic tasks that control top-down and bottom-up flows. Examples include: BRANCH, which selects an output path at runtime based on a boolean condition; JOIN, which merges multiple paths into one; FORK, which allows multiple concurrent strategy paths; REDUCE, which consolidates the results of multiple strategy paths into one; and STOP, which terminates the design flow;

\item \textbf{$O$-task}: These are self-contained optimizing tasks that enhance deep learning models based on specific objectives and constraints. Our current pipe task repository includes PRUNING and SCALING, which are implemented using the Keras API (version 2.9.0), and QUANTIZATION, which is performed using C++ source-to-source transformations via the Artisan framework~\cite{fccm20_artisan};

\item \textbf{$\lambda$-task}: These tasks perform functional transformations on the model space, such as compilation and synthesis. Examples include HLS4ML (version 0.6.0), which translates a DNN model into an HLS C++ model, and Vivado HLS (version 20.1), which translates an HLS C++ model into an RTL model.

\end{itemize}

It is worth noting that our framework is customizable, allowing users to create and integrate their own pipe tasks tailored to their specific needs.

\subsection{Building a Strategy}

\begin{lstlisting}[float=tp,emph={with},emphstyle={\bfseries}, language=Python, caption=The implementation of the pruning strategy using our framework., label=code:strategy]
# The pruning strategy architecture - Fig. 15(a)
from MetaML import * 
# design-flow architecture
with Dataflow() as df:
    join = Join() << KerasModelGen() 
    branch = Branch('B') << (VivadoHLS() << 
             (HLSML() << (Pruning() << join)))

    branch >> [join, Stop()]
# design-flow configuration
cfg = {
    'KerasModelGen::train_en': False,
    'Pruning::tolerate_accuracy_loss': 0.02,
    'Pruning::pruning_rate_threshold': 0.02,
    'B@fn': lambda metamodel:...
    'HLS4ML::default_precision': 'ap_fixed<18,8>',
    'HLS4ML::IOType': 'io_parallel',
    'HLS4ML::FPGA_part_number': 'xc7z020clg400-1',
    'HLS4ML::clock_period' : 10,
    'train_test_dataset': '/mypath/target_dataset',
    'train_epochs': 15,
    'Stop::fn': lambda metamodel: ... 
}
# run design-flow  
optimised_model = df.run(cfg)
\end{lstlisting}

Listing~\ref{code:strategy} provides the Python code for implementing the pruning strategy depicted in Fig.\ref{fig:design_flow}(a) and discussed in Section~\ref{sec:single_opt}. The code consists of three sections: \textbf{(i)}~lines 4--9 builds the design-flow architecture graph by connecting the corresponding tasks. The $>>$ and $<<$ operators are utilized to define the connections between task instances, and the default name of each task instance serves as its identifier, although it can be overridden by providing it as an argument (line 6). The end result is a cyclic-directed graph. \textbf{(ii)}~lines 11--23 define the configuration object, which holds all the parameters for configuring the tasks. Parameters referred to as \texttt{TaskType::parameter} are passed to all tasks of that task type (lines~16--19), whereas parameter names in the format of \texttt{InstanceName@parameter} are passed to a specific instance (line 15). All other parameters are global (lines 20--21) and accessible to any task.  Finally, \textbf{(iii}) line~25 executes the design-flow by passing the configuration object.

The execution of the design-flow starts with graph validation, which ensures that there is at least one source task and no multiplicity constraints are violated. Tasks are executed by an internal scheduler using a thread pool, which assigns jobs to available workers or stalls until a worker becomes available. After validating the architecture graph, an empty meta-model is generated using the supplied configuration, and the scheduler runs the source task. When a task completes execution, it submits a job to execute the next set of tasks in the flow. The STOP task terminates execution after all submitted jobs are completed. Modifying the design-flow architecture and updating its configuration parameters can lead to new optimization strategies, which is discussed next.

\newtext{Although this implementation may resemble a Python loop that sequentially invokes a series of functions, the \texttt{Dataflow} layer supports a fundamentally different architectural model to codify a design-flow. Rather than encoding a fixed linear script, it realizes a blackboard architecture~\cite{carver1994evolution}, where each $O$/$K$/$\lambda$-task is an independent and potentially heterogeneous component, operating at the software, HLS, or hardware-report level, that does not invoke other tasks directly. All interaction instead occurs through the shared meta-model, which acts as the blackboard and stores model versions across abstraction levels, configuration parameters, tool reports, metrics, and execution logs in a unified structure. A controller uses this evolving state to determine which task should run next, enabling behaviors such as parallel exploration (FORK), conditional re-routing based on hardware constraints (BRANCH), merging of alternative paths (REDUCE), and revisiting earlier stages when necessary. This design allows heterogeneous components to be composed without predefined interfaces and provides the minimal structure required for cross-stage, feedback-driven optimization, which cannot be captured by a simple loop-based implementation.}
\section{Implementation}
\label{sec:implementation}

\subsection{Co-Optimization Workflow}
\begin{figure}[tp]
    \begin{center}
    \includegraphics[width=0.8\linewidth]{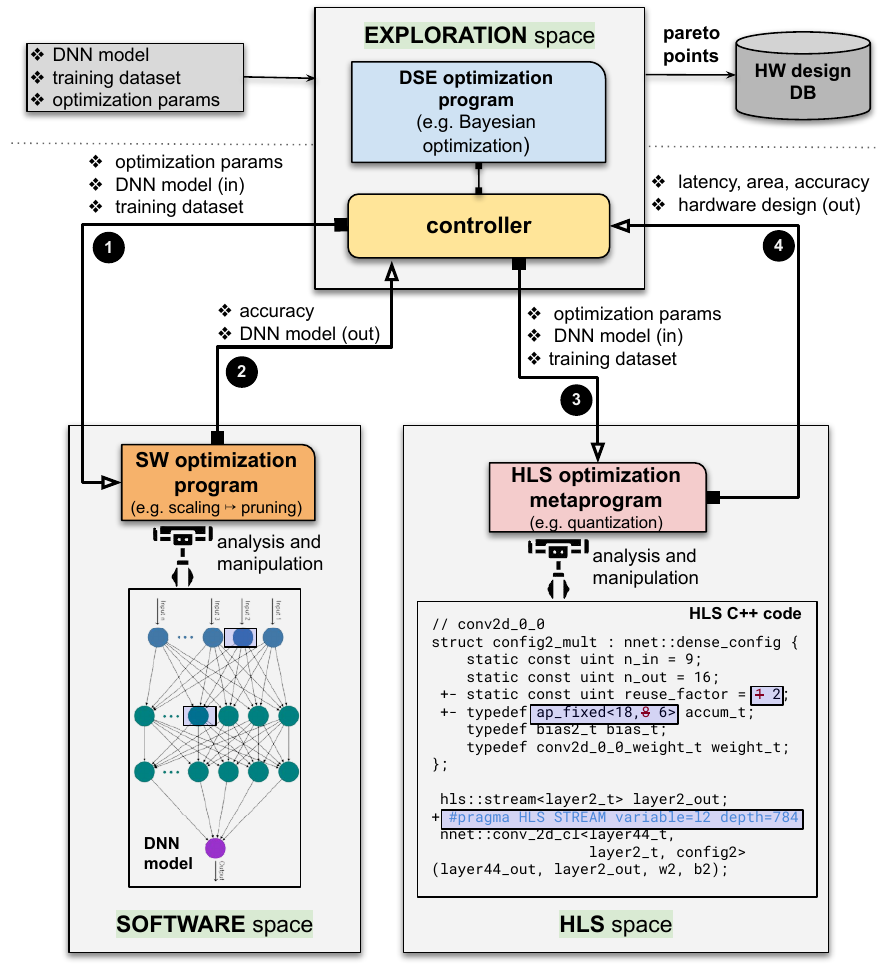}
    \end{center}
       \caption{This figure illustrates the implementation of our co-optimization framework, featuring an organized system of \textit{optimization} spaces, each autonomously running Python programs within dedicated environments, overseen by an \textit{exploration} space executing a general optimization strategy. The diagram depicts the orchestration of local optimization spaces, such as software and hardware, through a controller process.}
    \label{fig:opt_flow}
\end{figure}

As illustrated in Fig.~\ref{fig:opt_flow}, our co-optimization approach allows the encoding of optimizations across different models of computation - from software to hardware (see \textbf{C1} in Section~\ref{sec:intro}). At the top level, a DSE optimization program is employed (see \textbf{C2}), executing a comprehensive exploration strategy and overseeing local optimizations at two (or more) distinct levels of abstraction. At the software DNN level, optimizations are executed on the architecture using tools like Tensorflow or PyTorch. At the HLS DNN level, optimizations are conducted by modifying the HLS code automatically  using metaprogramming. Optimizing DNN models across multiple abstraction levels, from software to hardware, leverages complementary strategies that address different bottlenecks within the system. Software optimizations prioritize algorithmic efficiency and reducing computational complexity, while hardware optimizations target improvements in throughput, latency, resource utilization and energy efficiency.

Our co-optimization workflow begins with the Exploration space. Here, a DSE program utilizes a DNN model, training data, and configuration parameters to conduct the exploration workflow. This process yields a variety of hardware designs with trade-offs in metrics such as performance and accuracy. A controller process orchestrates this exploration by managing communication between the DSE program and local optimization spaces. Prior to exploration, the DSE program sets up and deploys optimization programs within these spaces. Fig.~\ref{fig:opt_flow} depicts two such spaces: one for software (SW) utilizing TensorFlow and another for HLS employing Vivado HLS. Each space optimizes the model according to selected strategies, such as pruning or quantization.

After system setup, the exploration loop commences. In each iteration: \circled{1} The DSE program dispatches the DNN model and a specific parameter set to the software optimization space, where tasks like scaling and pruning are executed. \circled{2} Upon optimization completion, the refined model is returned to the controller along with its associated accuracy. \circled{3} Subsequently, the model is forwarded to the HLS space for additional optimization. A metaprogram adjusts the source code based on recommended optimizations (e.g., quantization) and parameter values from the controller. After synthesis, metrics such as power, area, and accuracy are obtained and \circled{4} fed back to the Exploration space. The DSE program evaluates designs using these metrics, retains promising candidates, and suggests new parameters for the next iteration.

This separation of concerns into distinct optimization spaces—software, HLS, and DSE—allows for the integration of diverse optimization strategies and techniques. It enables the introduction of faster or more effective exploration methods without altering existing optimizations. In addition, it combines exploration across software and hardware domains to identify effective techniques for optimizing performance. Moreover, it facilitates specialization of new hardware optimizations specific to certain FPGA HLS tools by allowing HLS code to be customized through metaprogramming.

\subsection{Auto-Pruning with Binary Search}

Pruning is a technique that improves the performance and efficiency of neural networks by removing insignificant weights. The pruning strategy uses a PRUNING $O$-task, which gradually zeroes out weights during training to create a more compact and efficient network while maintaining accuracy. This optimization task supports auto-pruning, which automatically determines the highest pruning rate while maintaining a given level of accuracy loss. The PRUNING $O$-task is illustrated in Fig.~\ref{fig:design_flow}(a). Formally, the objective of this $O$-task is defined as:
\begin{equation}
\begin{footnotesize}
\begin{aligned}
& \underset{}{\text{maximum}}
& & Pruning\_rate \\
& \text{subject to}
& & Accuracy\_loss(Pruning\_rate) \leq \alpha_p 
\end{aligned}
\end{footnotesize}
\end{equation}

Starting at 0\% pruning rate, the auto-pruning algorithm obtains initial accuracy
$Acc_{p0}$ at step~1 (s1).
It then uses a binary search approach, increasing or decreasing the pruning rate based on whether the accuracy loss is within a user-defined tolerance ($\leq \alpha_{p}$) as shown in Table~\ref{table:pipetasks}. The algorithm terminates when the rate difference is below a threshold ($\beta_{p}$). The number of steps is determined by $1+log_{2}(1/\beta_{p})$. 

\subsection{Auto-Scaling} 

To accommodate large DNN designs on FPGAs, we adopt the SCALING $O$-task that automatically reduces the layer size while monitoring the result accuracy loss, denoted by $\alpha_s$. 
The objective of this $O$-task is formally defined as:
\begin{equation}
\begin{footnotesize}
\begin{aligned}
& \underset{}{\text{minimum}}
& & Scaling\_factor \\
& \text{subject to}
& & Accuracy\_loss(Scaling\_factor) \leq \alpha_s 
\end{aligned}
\end{footnotesize}
\end{equation}

The auto-scaling process begins with a scaling factor of 1 (i.e., no scaling), getting initial accuracy
$Acc_{s0}$ at step~1 (s1). The algorithm then gradually reduces the scaling factor, continuing the search until the accuracy loss exceeds the threshold $\alpha_s$.  
If necessary, $\alpha_s$ can be adjusted to achieve further size reduction with minimal impact on accuracy.

\subsection{Auto-Quantization  with Heuristic Search}
\label{sec:qhs}
In our framework, we introduce quantization task based on the Quantization Heuristic Search (QHS) algorithm~\cite{que2024optimizing}, a post-quantization method. It automatically determines the lowest bitwidth for each layer while maintaining a given level of accuracy loss. Formally, the objective of this task is defined as:
\begin{equation}
\begin{footnotesize}
\begin{aligned}
& \underset{}{\text{maximum}}
& & Bit\_width\_reduction \\
& \text{subject to}
& & Accuracy\_loss(Bit\_width\_reduction) \leq \alpha_q 
\end{aligned}
\end{footnotesize}
\end{equation}
It performs mixed-precision quantization for DNNs, allowing separate precisions for the weights, biases and results of layers of a model, taking the advantages of the customizability of FPGAs. 
While this customizability creates a massive search space of quantization configurations, dependencies within the network can be used to reduce the degrees of freedom of the design space.

\begin{figure}
\begin{center}
\includegraphics[width=0.80\linewidth]{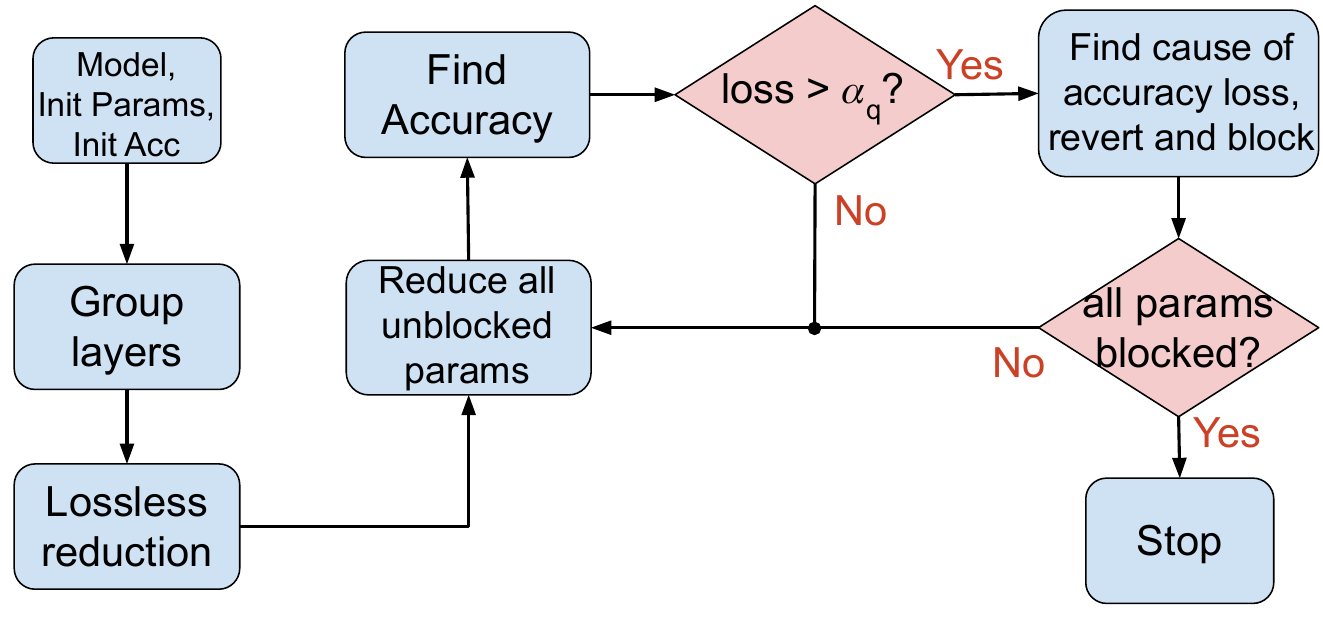}
\end{center}
   \caption{The proposed QHS algorithm}
\label{fig:quantization_algo}
\end{figure}

Fig.~\ref{fig:quantization_algo} shows the steps of the proposed QHS algorithm. The optimization task begins by accepting a model with a maximum tolerable accuracy loss, denoted as $\alpha_q$. Initially, virtual layers are created, and lossless reduction is performed by minimizing the integer bits representing model parameters using the base-2 logarithm of the largest weight and bias in each layer, plus one bit for the sign to avoid saturation. Subsequently, all parameters (weights, biases and results) are initially assumed to be further reducible. The total bit-widths of these parameters are then reduced by one bit if they are considered reducible. This quantization is applied to the C++ kernel, and the accuracy is evaluated through a runtime simulation. If the accuracy loss remains within the user-specified tolerance, $\alpha_q$, this step is repeated. If the accuracy loss exceeds the tolerance, the algorithm identifies and blocks non-reducible precisions. It determines which precisions are most sensitive by testing the reduction impact on accuracy. If a bit-width reduction breaks the accuracy constraint, that precision is marked non-reducible. The process repeats until no further reductions are possible within the accuracy constraints. Support for Quantization-aware training, such as HGQ~\cite{sun2026hgq}, is left to future work.

\subsection{Metaprogramming for HLS Design Optimization}\label{sec:metaprogramming}

\begin{figure}[tp]
    \begin{center}
    \includegraphics[width=0.8\linewidth]{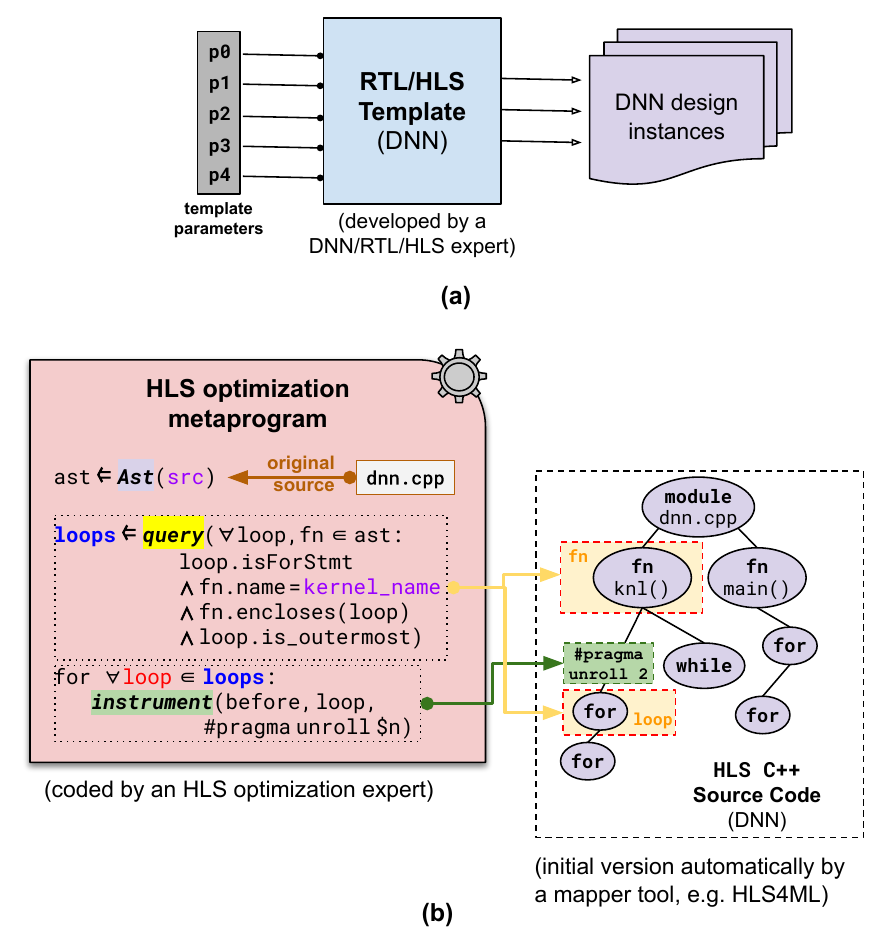}
    \end{center}
       \caption{ (a) Current methods for deriving hardware designs rely on templates developed by experts. However, these templates are constrained in generating a finite number of design variants from predetermined parameter values and demand expertise for accommodating new DNN architectures.
(b) An HLS metaprogram in pseudocode illustrating dynamic C++ code manipulation through Abstract Syntax Tree construction to derive an optimized design variant.}
    \label{fig:metaprogramming}
\end{figure}

A novel aspect of our co-optimization approach is the integration of metaprogramming into hardware design optimization, \textcolor{\mycolor}{which enables the implementation of the Quantization Heuristic Search (QHS) algorithm described in Section~\ref{sec:qhs}}. In this section, we describe the key motivation and benefits of metaprogramming, and how we integrate it in our co-optimization framework.

\begin{color}{\mycolor}
\textbf{\textit{Beyond template-based and code generator approaches.}}  Template-based frameworks and code generator modifications (for instance, extending HLS4ML) are effective when optimization choices can be captured as a fixed set of global parameters. For example, templates can expose settings such as default precision, loop unroll factors, or hardware interface options. Similarly, modifying a code generator allows these parameters to be realized directly in generated code. However, both approaches share two fundamental limitations:

\begin{itemize}
\item \textbf{Restricted scope of changes:} Templates and generators can only express transformations that were foreseen and encoded by their designers. Adding new optimisation dimensions requires extending the generator itself, which is costly and often tool-specific.

\item \textbf{Limited applicability across sources:} In practice, real workflows mix generated code (e.g., HLS4ML) with user-written kernels, vendor libraries, and legacy C++. Template extensions or generator patches cannot easily propagate optimisations across such heterogeneous sources.
\end{itemize}

\textit{\textbf{Benefits of metaprogramming.}} Our metaprogramming approach addresses these limitations by operating directly on the C++ Abstract Syntax Tree (AST). This enables:

\begin{itemize}
\item Local, structural optimizations such as loop-specific pragma placement, which templates cannot expose without proliferating knobs.

\item Cross-kernel consistency edits, for example rewriting all types in a conv → pool → batch norm cascade to ensure coherent quantisation without re-generation.

\item Device-aware refactorings guided by design-space exploration, where metaprograms inject vendor-specific constructs based on feedback.

\end{itemize}

Quantization illustrates this distinction clearly. A template or generator can reduce precision globally (e.g., set all layers to 8-bit), but cannot ensure type consistency across intermediate streams or flexibly adapt precision to device constraints without modifying the generator itself. Our metaprogram, by contrast, queries the AST to identify cascaded kernels, rewrites typedefs and template arguments in one pass, and produces a coherent low-precision dataflow candidate automatically.

In summary, templates and generators are well-suited for global configuration, while metaprogramming is essential when optimisation requires local, structural, or device-specific edits. This generality allows the same refactoring (e.g., quantisation) to be applied not only to HLS4ML output, but equally to handwritten or third-party C++ code, demonstrating the broader utility of our approach.
\end{color}

\textit{\textbf{Example of an HLS metaprogram.}} Fig.~\ref{fig:metaprogramming}(b) illustrates an HLS metaprogram that dynamically modifies a given C++ source file using Python. The process begins with the construction of an Abstract Syntax Tree (AST) representation of the C++ code. The AST provides a structured representation of the source code, allowing precise modifications for any of its constructs. The metaprogram then performs code querying and analysis to identify outermost loop structures within kernel functions. These loops are critical for performance optimization, as their execution directly impacts resource utilization and latency.

Once the relevant loops are identified, the metaprogram instruments the code by inserting pragma directives such as \texttt{\#pragma unroll \$n}. These directives control loop unrolling dynamically, enabling optimized parallel execution of computations. By automating this process, the metaprogram allows rapid exploration of different optimization strategies without requiring manual code modifications. This automation significantly improves both efficiency and design flexibility, making it easier to adapt to different DNN architectures and FPGA platforms. A more detailed explanation
about querying and instrumentation capabilities can be found in
~\cite{fccm20_artisan}.

\begin{figure}[tp]
    \begin{center}
    \includegraphics[width=0.75\linewidth]{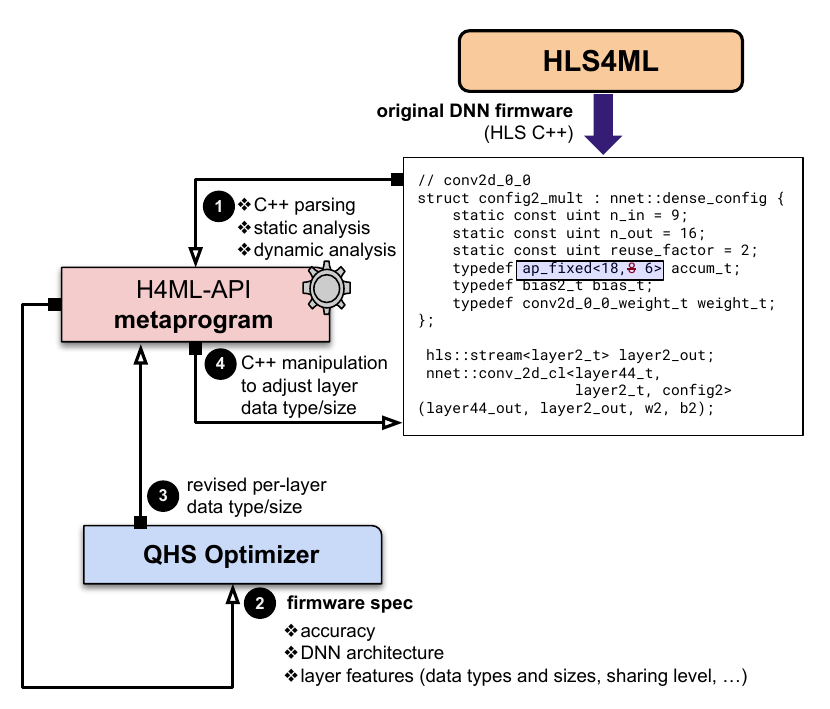}
    \end{center}
       \caption{ Integration of the Quantization Heuristic Search (QHS) algorithm with HLS4ML and metaprogramming. The metaprogram dynamically adjusts data types and bit-widths at the HLS C++ level, optimizing computational efficiency while maintaining model accuracy. }
    \label{fig:metaprogramming_quant}
\end{figure}

\textit{\textbf{Application in quantization optimization.} }
We integrate Quantization Heuristic Search (QHS), detailed in Section~\ref{sec:qhs}, as an optimization task within our framework using metaprogramming to refine fixed-point numerical representations directly at the HLS C++ level. As illustrated in Fig.~\ref{fig:metaprogramming_quant}, we develop a metaprogram called H4ML-API, which programmatically analyzes and modifies HLS4ML-generated code to automate the quantization process.

The workflow proceeds as follows. First, the HLS4ML tool generates an initial, unoptimized firmware from a DNN model. Then, \circled{1}~the H4ML-API metaprogram inspects the source structure and executes the firmware to extract DNN architecture details, layer characteristics, and accuracy metrics. This information is \circled{2}~passed to the QHS process, which identifies opportunities for bit-width reduction while maintaining model accuracy, dynamically adjusting bit-widths based on layer-specific requirements. The refined bit-width settings are then \circled{3}~transferred back to the H4ML-API, which \circled{4}~modifies the HLS C++ code accordingly. This iterative process continues from step \circled{1} while the accuracy loss remains within the $\alpha_q$ threshold.

A key feature of this approach is the creation of virtual layers to optimize bit-width allocation. Rather than treating each operation separately, the metaprogram groups cascaded operations into logical units, enabling more effective quantization. For instance, if a convolutional layer produces an 8-bit output, subsequent operations such as max-pooling and batch normalization are assigned the same precision, avoiding redundant higher-bit computations. By systematically reducing bit-widths without compromising accuracy, this approach significantly decreases resource usage while maintaining high inference performance.

\subsection{Design Space Exploration (DSE)}

\begin{algorithm}
\small
\caption{\textcolor{\mycolor}{MetaML-Pro: Cross-Stage Constrained BO with Bottom-Up Feedback and Order Search}}
\label{alg:metamlpro-dse}
\DontPrintSemicolon
\SetKwProg{Fn}{Function}{}{}
\SetKw{KwAnd}{and}
\SetKw{KwOr}{or}
\SetKw{KwNot}{not}

\textcolor{\mycolor}{
\KwIn{
  Optimization tasks $O=\{\text{SCALING (S)},\text{PRUNING (P)},\text{QUANTIZATION (Q)}\}$; \\
  Orderings $\Pi \subseteq \text{Perm}(O)$ (e.g., $\{S{\to}P{\to}Q,\;P{\to}S{\to}Q,\dots\}$); \\
  Initial parameters $\theta_1=\tau_1=(\alpha_p,\alpha_s,\alpha_q)$; \\ 
  Resource/latency/accuracy limits $u_{\max}$, $t_{\max}$ $acc\_loss_{\max}$; \\
  BO budget $T$; initial design set $\mathcal{D}_0$; vendor set $\mathcal{V}=\{\text{AMD},\text{Intel}\}$; \\
  Meta-model $MM$ with sections CFG/LOG/MODEL-SPACE.
}
\KwOut{Best design(s) $\mathcal{D}^\star$ and Pareto set $\mathcal{P}$.}
\BlankLine
\textbf{Initialize:} 
Fit surrogate $\mathcal{S}$ on $\mathcal{D}_0$; 
Set BO results $\mathcal{E}_\pi \gets \emptyset$ for all $\pi \in \Pi$.\;
\For(\tcp*[f]{FORK over orderings}){$\pi \in \Pi$ \textbf{in parallel}}{
  \For{$t \gets 1$ \KwTo $T$}{
    $m_t \gets \textsf{Evaluate}(\pi,\theta_t,MM)$\;
    \If{$\textsf{Feasible}(m_t,\theta_t,u_{\max},t_{\max}, acc\_loss_{\max})$}{
      $s_t \gets \textsf{Score}(\textsf{Normalize}(m_t))$\;
    }
    \Else{
      $s_t \gets -M$\tcp*{constraint-violated candidate receives large negative score}
    }
    $\mathcal{S} \gets \textsf{UpdateSurrogate}(\mathcal{S}, \theta_t, s_t)$\tcp*{Fit surrogate}
    $\mathcal{E}_\pi \gets \mathcal{E}_\pi \cup \{(\theta_t, m_t)\}$ \;
    $\theta_{t+1} \gets \textsf{ProposeCandidate}(\mathcal{S}, \text{CFG}(MM), \pi)$\tcp*{acquisition-guided}
  }
}
$\mathcal{P} \gets \textsf{ParetoReduce}\!\left(\bigcup_{\pi \in \Pi} \mathcal{E}_\pi\right)$ \tcp*{REDUCE across paths}
$\mathcal{D}^\star \gets \textsf{SelectBest}(\mathcal{P}, \text{policy})$\;
}
\textcolor{\mycolor}{
\Return $(\mathcal{D}^\star, \mathcal{P})$\;
\BlankLine
\Fn{\textsf{Evaluate}$(\pi,\theta,MM)$}{
  \tcp{Top-down pass with inner DSE inside each $O$-task}
  $\textsf{Apply}(MM.\text{CFG}, \theta)$\;
  $\textsf{Run}(\text{KERAS-MODEL-GEN})$; \\
  \ForEach{$O \in \pi$}{
    $\textsf{Run}(O)$\tcp*{SCALING/PRUNING/QUANT.: each has its own inner DSE}
  }
  $\textsf{Run}(\text{HLS4ML})$\;
  \tcp{Vendor-aware BRANCH}
  \eIf{$\textsf{VendorSelect}(MM) = \text{AMD}$}{
    $\textsf{Run}(\text{VIVADO-HLS})$
  }{
    $\textsf{Run}(\text{INTEL-oneAPI})$
  }
  $m \gets \textsf{CollectMetrics}(\text{accuracy}, \text{latency}, \text{DSP}, \text{LUT}, \text{FF}, \text{BRAM})$\;
  \Return $m$\;
}
\Fn{\textsf{Feasible}$(m,\theta,u_{\max}, t_{\max}, acc\_loss_{\max})$}{
  \Return $\big(m.\text{util} \le u_{\max}\big)\;\KwAnd\; \big(m.\text{lal} \le t_{\max}\big)\; \KwAnd\; \big(m.\text{acc\_loss} \le acc\_loss_{\max}\big)$\;
}
\Fn{\textsf{Score}$(\tilde{m})$)}{
  \Return $\;-\; (w_1\,\tilde{m}.\text{Acc} \;+\; w_2\,\tilde{m}.\text{DSP} \;+\; w_3\,\tilde{m}.\text{LUT}\;+\; w_4\,\tilde{m}.\text{lat})$\;
}
}
\end{algorithm}

Within our approach, Design Space Exploration (DSE) algorithms are critical for automatically adjusting the optimization parameters in subsequent iterations to improve design performance, while maintaining high model accuracy and meeting the constraints of the device. 
\textcolor{\mycolor}{
We use Bayesian optimization to select high-level configuration parameters $\boldsymbol{\theta}$ for the cross-stage flow, e.g., global tolerances $\boldsymbol{\tau}=(\alpha_p,\alpha_s,\alpha_q)$. Layer-wise details are decided inside each $O$-task’s inner search (Sections~4.2–4.4), while BO governs the outer loop (Algorithm~\ref{alg:metamlpro-dse}).
}

\textcolor{\mycolor}{
\subsubsection{\textbf{Surrogate model.}}
We model the scalarized objective $f(\boldsymbol{\theta})$ with a zero-mean Gaussian Process (GP) using a Mat\'ern covariance function and homoscedastic Gaussian noise. Inputs are min-max normalized to $[0,1]$. GP hyperparameters are re-estimated by maximizing the marginal likelihood at each iteration using the current design set.
}

\textcolor{\mycolor}{
\subsubsection{\textbf{Objective scalarization and normalization.}}
Let results be $\text{Acc}, \text{DSP}, \text{LUT}, \text{Lat}$. We normalize each metric to $[0,1]$ using ranges observed so far, then form the minimization objective
\[
f(\boldsymbol{\theta}) \;=\;-(\,w_1\,\tilde{\text{Acc}} \;+\; w_2\,\tilde{\text{DSP}} \;+\; w_3\,\tilde{\text{LUT}} \;+\; w_4\,\tilde{\text{Lat}}) \;
\]
subject to feasibility. Unless otherwise stated we use equal weights. To target particular frontiers we choose weights accordingly; e.g., Accuracy-DSP uses $(w_1,w_2,w_3,w_4)=(0.5,0.5,0,0)$, Accuracy-LUT uses $(0.5,0,0.5,0)$, and the balanced Acc-DSP-LUT-Lat setting uses $(0.25,0.25,0.25,0.25)$.
}

\textcolor{\mycolor}{
\subsubsection{\textbf{Constraints.}}
Feasibility requires: (i) device utilization within limits (DSP/LUT), (ii) latency $\le t_{\max}$, and (iii) model accuracy loss $\le Acc\_loss_{max}$. We treat constraints as \emph{hard}: infeasible candidates are rejected and scored as $-M$ in Algorithm~\ref{alg:metamlpro-dse}’s \texttt{Score} function. 
}

\subsubsection{\textbf{Acquisition.}}
The Bayesian optimization algorithm finds next candidate in the search space by maximizing the acquisition function's value, where the acquisition function is Expected Improvement (EI).

\subsubsection{\textbf{Convergence.}}
The outer BO loop stops when any of the following holds: (i) budget $T=22$ iterations per ordering $\pi$, or (ii) the incumbent design remains unchanged for 5 iterations. Inner loops terminate as defined earlier: pruning uses a binary search until $|\Delta \text{rate}|<\beta_p$ (Section~4.2), scaling stops when loss exceeds $\alpha_s$ (Section~4.3), and QHS halts when no further reducible precisions remain under $\alpha_q$ (Section~4.4).

\subsubsection{\textbf{Global tolerance variables.}}
We treat the vector of global tolerances $\tau=(\alpha_p,\alpha_s,\alpha_q)$ as part of the BO decision variable $\boldsymbol{\theta}$. The user supplies only an initial $\tau_0$ and bounds $[\tau_{\min},\tau_{\max}]$. At each BO iteration, the acquisition proposes a new $\boldsymbol{\theta}_t$ that includes $\tau_t$; no per-benchmark manual tuning of $\alpha_{x}$ is required. Inner searches remain local to each $O$-task, while BO governs the outer loop over $\tau$ and flow-level choices (Algorithm~\ref{alg:metamlpro-dse}).%

\subsection{DSE-Generated Hardware Architecture}
We adopt a dataflow-style, layer-pipelined streaming hardware architecture, which is well-suited for implementing DNN designs. All candidates explored by MetaML-Pro are instantiated from the \texttt{hls4ml} hardware library as per-layer compute kernels (e.g., \texttt{nnet::conv\_2d\_cl}, and \texttt{nnet::dense}) connected by streaming channels. Concurrency (e.g., unrolling/reuse, initiation interval) and buffering (e.g., FIFO depths) are driven by HLS pragmas and layer configuration parameters. The library is initially configured via \texttt{IOType}, \texttt{default\_precision}, \texttt{device}, and \texttt{clock} (Table~\ref{table:pipetasks}). In addition, the generated HLS can be refined by metaprogramming that inserts/tunes pragmas and C++ codes through AST manipulation (Section~\ref{sec:metaprogramming}).

\begin{figure} 
   \centering
  \subfloat[\label{fig:folded}]{%
    \includegraphics[width=0.40\linewidth]{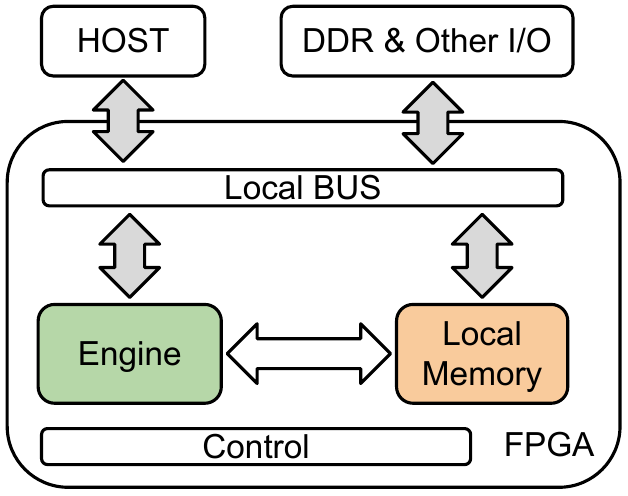}}
    \hspace*{\fill}
  \subfloat[\label{fig:partialfolded}]{%
    \includegraphics[width=0.52\linewidth]{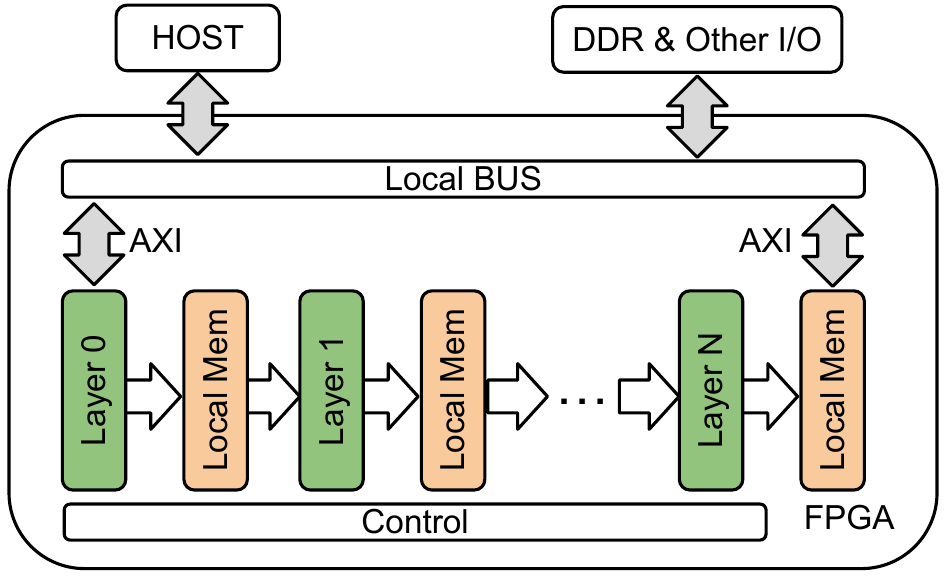}}
    
  \caption{\newtext{FPGA-based DNN accelerator architectures: (a) A single-engine hardware architecture with a shared computation engine that is time-multiplexed across all layers; (b) A stream-based layer-wise pipeline hardware architecture with several custom engines, each processing a sub-layer or a whole layer or even multiple layers in a DNN model.}
  }
  \label{fig:architectures} 
\end{figure}

This architecture adopts a layer-wise pipeline~\cite{nakahara2020high, que2022optimizing,  que2024ll, sun2025fast}, as shown in Fig.~\ref{fig:architectures}(b). Each major computational block of DNN models is mapped onto a tailored hardware module, with local memory connected via on-chip streams. The modules are chained into a deep pipeline so that different layers process different inputs concurrently, enabling continuous dataflow. In contrast to a conventional single-engine design~\cite{que2020optimizing, que2021recurrent, fan2021TNNLS} that time-multiplexes a common compute core across all layers (Fig.~\ref{fig:architectures}(a)), this architecture allows each block in DNN to be independently optimized for resource utilization, numerical precision, and micro-architecture. 
Intermediate data remain on-chip between layers to avoid repeated off-chip memory accesses and reduce inference latency, while exploiting the full potential of FPGA customizability. Please refer to~\cite{que2023reconfigurable} for a more detailed discussion of these two architectures.

In regards to the DSE process, our proposed Bayesian-optimization (BO) loop treats the global tolerance vector $\tau=(\alpha_p,\alpha_s,\alpha_q)$ as part of the decision variable and \emph{automatically} proposes new $\tau_t$ each iteration. For each BO proposal, the $O$-tasks run inner searches that directly modify the hardware: (i) Pruning deletes weights/filters, shrinking convolutions and dense layers; (ii) Scaling reduces layer sizes; (iii) \textsc{Quantization} (QHS) assigns mixed-precision formats per virtual layer at the HLS C++ level. The outcome is an HLS-based C++ design with a specific level of parallelism, buffering, and custom numeric formats, ready for synthesis and evaluation.

\subsection{Collaborative Optimization Mechanics}
MetaML-Pro realizes collaboration across \emph{four interacting layers} spanning software and HLS domains. First, it employs a data/state-level collaboration via the meta-model. All tasks (software and HLS) read/write a shared meta-model (CFG/LOG/MODEL-SPACE), which carries models, tool reports, and computed metrics across stages. This allows later-stage hardware outcomes (latency/utilization) to inform earlier software optimizations programmatically. Second, it employs control-level collaboration via $K$-tasks. BRANCH triggers bottom-up adaptation when constraints are violated; FORK/REDUCE explore alternative orderings and merge by Pareto dominance; JOIN composes multi-stage flows. This moves beyond static pipelines to \emph{runtime} flow reconfiguration. Third, it has parameter-level collaboration via Bayesian Optimization based outer loop and inner local searches. The outer constrained BO jointly chooses the global tolerance vector $\tau=(\alpha_p,\alpha_s,\alpha_q)$ and searches across task orderings $\Pi$; each O-task then runs its own inner DSE (binary-search pruning, progressive scaling, QHS quantization). This couples global, cross-stage choices with local decisions automatically. 
And finally, the HLS metaprogram edits C++ kernels (AST-driven insertion/tuning, data-type/reuse updates) and applies QHS with virtual layers, propagating precision constraints across cascaded operations.

\begin{table}[tp]
\centering

\caption{Benchmarks used in this work}
\label{table:benchmark}

\scalebox{1.00}{
\begin{threeparttable}
\centering
\begin{tabular}{c|c|c}
\toprule
Model Name & Dataset & Application Domain \\ \midrule

Jet-DNN~\cite{duarte2018fast, coelho2021automatic} / Jet-CNN  & Jet-HLF~\cite{duarte2018fast}  & 
Jet Identification (Particle Physics)
\\ \midrule

VGG-7~\cite{tridgell2019unrolling, simonyan2014very} & MNIST~\cite{lecun1998gradient}  & Image Classification \\ \midrule

ResNet-9~\cite{hamanaka2023exploration, he2016deep}  & SVHN~\cite{netzer2011reading}  & Image Classification \\ \midrule

LSTM~\cite{que2020optimizing}  & MNIST  & Sequence classification \\ \midrule


\end{tabular}
\end{threeparttable} 
}
\end{table} 

\section{Evaluation}\label{sec:evaluation}
In this section, we demonstrate how optimization strategies can be built by revising design-flow architectures, combining and reusing pipe tasks, and modifying their configuration (Sections~\ref{sec:single_opt} and~\ref{sec:multi_opt}). We explain flow control in Sections~\ref{sec:bottom_up} and~\ref{sec:parallel_flow}. We discuss optimisation search strategies and compare evaluation results with other approaches in Sections~\ref{sec:opt_search} and~\secref{sec:discussion}, respectively.

\subsection{Experimental Setup}

Experiments were conducted in Python 3.9.15 with benchmark workloads from typical DNN applications, as presented in Table~\ref{table:benchmark}), including jet identification~\cite{duarte2018fast, moreno2020jedi}, image classification using VGG\cite{simonyan2014very} and ResNet~\cite{he2016deep} networks. 
The jet identification \textcolor{\mycolor}{(Jet-DNN)} task targeted FPGA-based CERN Large Hadron Collider (LHC) triggers with a 40 MHz input rate and a response latency of less than 1 microsecond. \textcolor{\mycolor}{Beyond Jet-DNN, we also evaluate VGG-7 (convolutional, MNIST), ResNet-9 (residual, SVHN), and LSTM (recurrent, sequential MNIST). These models are evaluated across single O-task strategies (pruning, quantization) and combined cross-stage strategies (e.g., S$\rightarrow$P$\rightarrow$Q), and we demonstrate portability across vendor toolchains (AMD/Xilinx and Intel). This breadth evidences that MetaML-Pro’s O-task library and K-task control (FORK/BRANCH/REDUCE) generalize across CNN, residual, and RNN topologies and across vendor flows.}
Default frequencies were 100MHz for Zynq 7020 and 200MHz for Alveo U250 and VU9P, and the HLS4ML task used 18-bit fixed-point precision with 8 integer bits.

\textcolor{\mycolor}{
Although our quantitative evaluation uses Jet-DNN, VGG-7, ResNet-9, and LSTM, MetaML-Pro’s compute is governed by its controller and task structure rather than these particular networks. Concretely, the outer Bayesian Optimization (BO) loop runs with a fixed iteration budget (the number of iteration and orderings by default) and convergence criteria that are agnostic to network depth/width (Algorithm.~\ref{alg:metamlpro-dse}), while each inner $O$-task scales gently with model size. 
Parallel FORK/REDUCE exploration and a thread-pool scheduler further amortize cost across orderings, and the meta-model caches intermediate artifacts across stages to avoid recomputation. These properties allow MetaML-Pro to scale to larger CNN and RNN models.  \newtext{We aim to extend our approach to emerging Transformer families in future work by plugging in the appropriate $\lambda$-tasks and kernels}.
}

\begin{figure} 
   \centering
   \hspace*{\fill}
  \subfloat[]{%
    \includegraphics[width=0.30\linewidth]{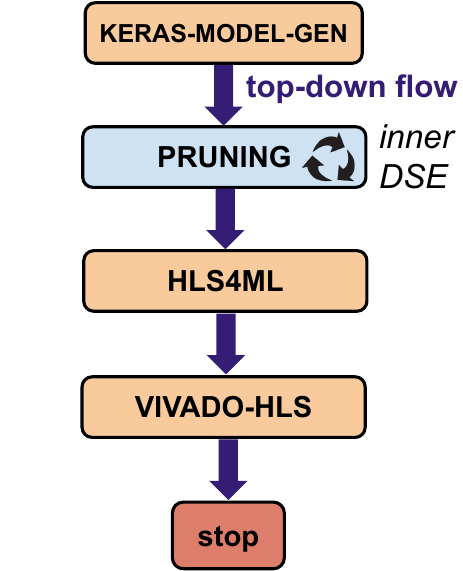}}
   \subfloat[]{%
    \includegraphics[width=0.32\linewidth]{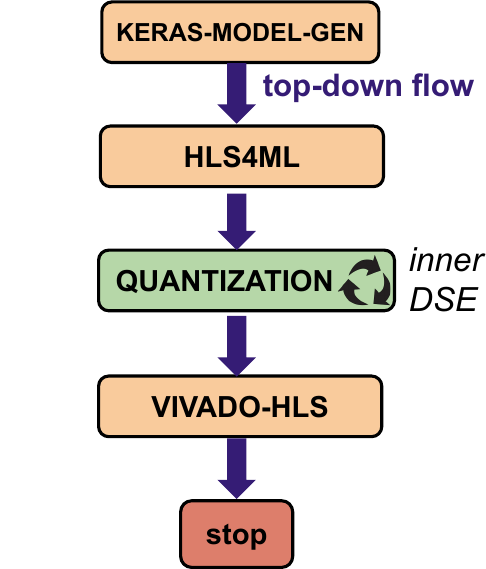}}
  \hspace*{\fill}
   \subfloat[]{%
    \includegraphics[width=0.33\linewidth]{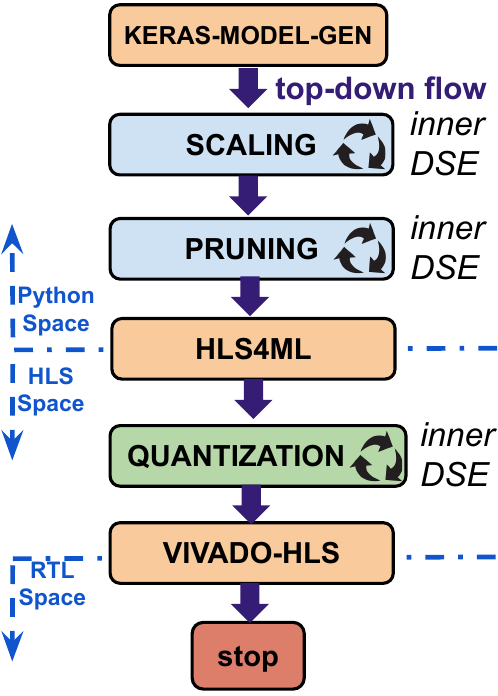}} 
 \hspace*{\fill}
  \caption{(a) Pruning strategy. (b) Quantization strategy. (c) The combined strategy of scaling, pruning and quantization. }
  \label{fig:design_flow} 
\end{figure}

\subsection{Single $O$-task Strategies}\label{sec:single_opt}

In this subsection, we focus on three strategies which are backed by a single $O$-task each, respectively. 

 \begin{figure} 
    \centering
   \subfloat[Jet-DNN\label{fig:lhc_dnn_pruning}]{%
     \includegraphics[width=0.49\linewidth]{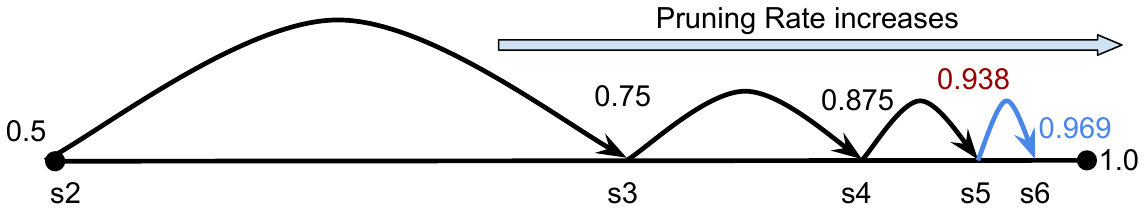}}
    \hspace*{\fill}
   \subfloat[Jet-CNN\label{fig:lhc_cnn_pruning}]{%
     \includegraphics[width=0.49\linewidth]{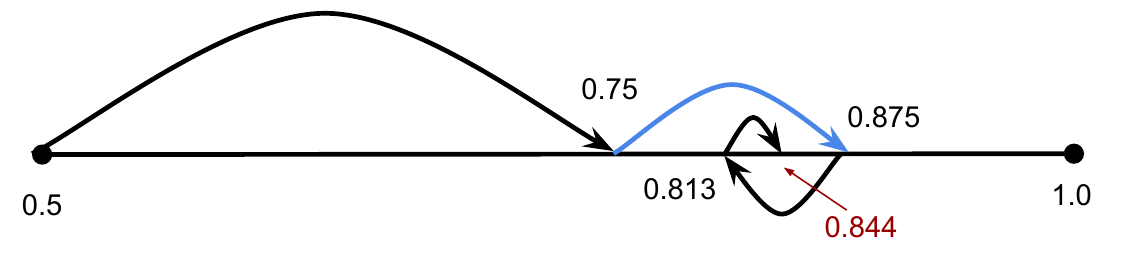}} 
   \\
   \subfloat[VGG7\label{fig:vgg7_pruning}]{%
     \includegraphics[width=0.49\linewidth]{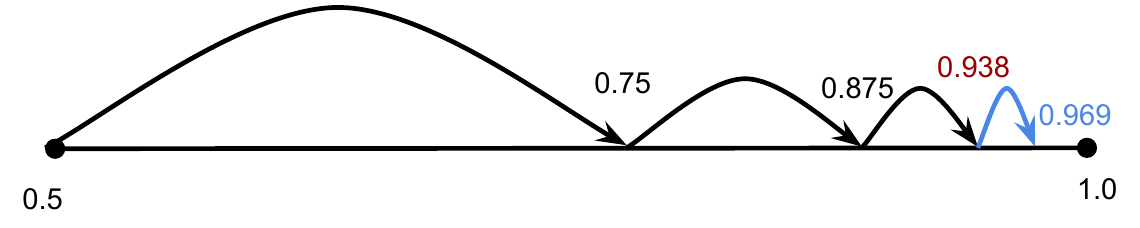}}
 \hspace*{\fill}
   \subfloat[ResNet9\label{fig:resnet8_pruning}]{%
     \includegraphics[width=0.49\linewidth]{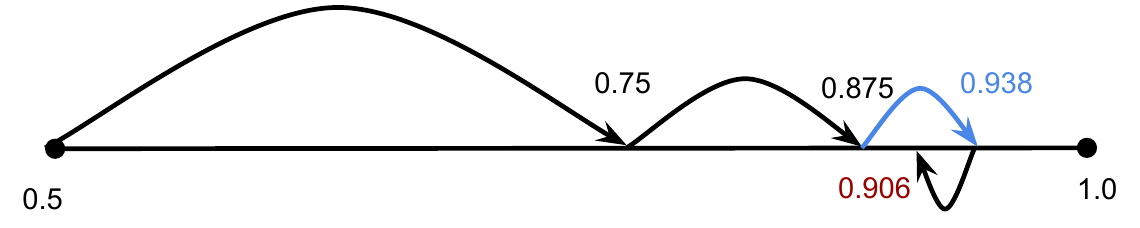}}
   \caption{ The auto-pruning algorithm applied to models with binary search direction shown. Omitting step s1 for visibility. The blue arrow indicates an accuracy loss $>$ user threshold; red denotes the optimal pruning rate. }
   \label{fig:auto_pruning} 
 \end{figure}

\begin{figure*} 
   \centering
  \subfloat[]{%
    \includegraphics[width=0.32\linewidth]{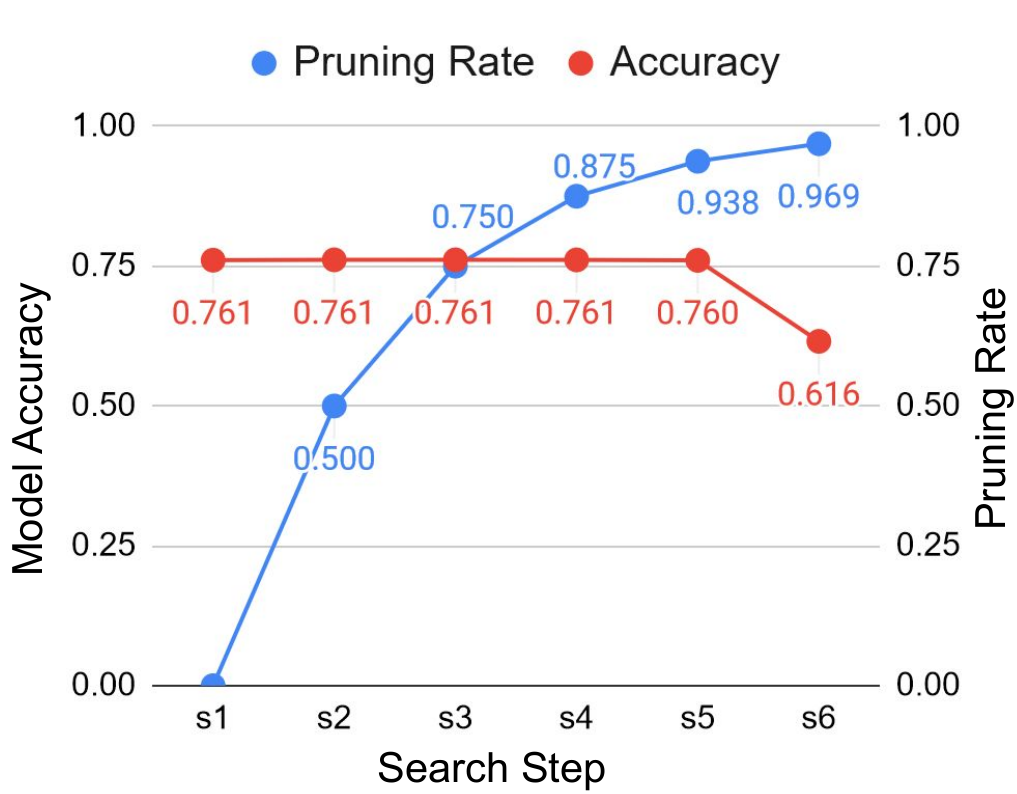}} 
   \hspace*{\fill}
  \subfloat[]{%
    \includegraphics[width=0.32\linewidth]{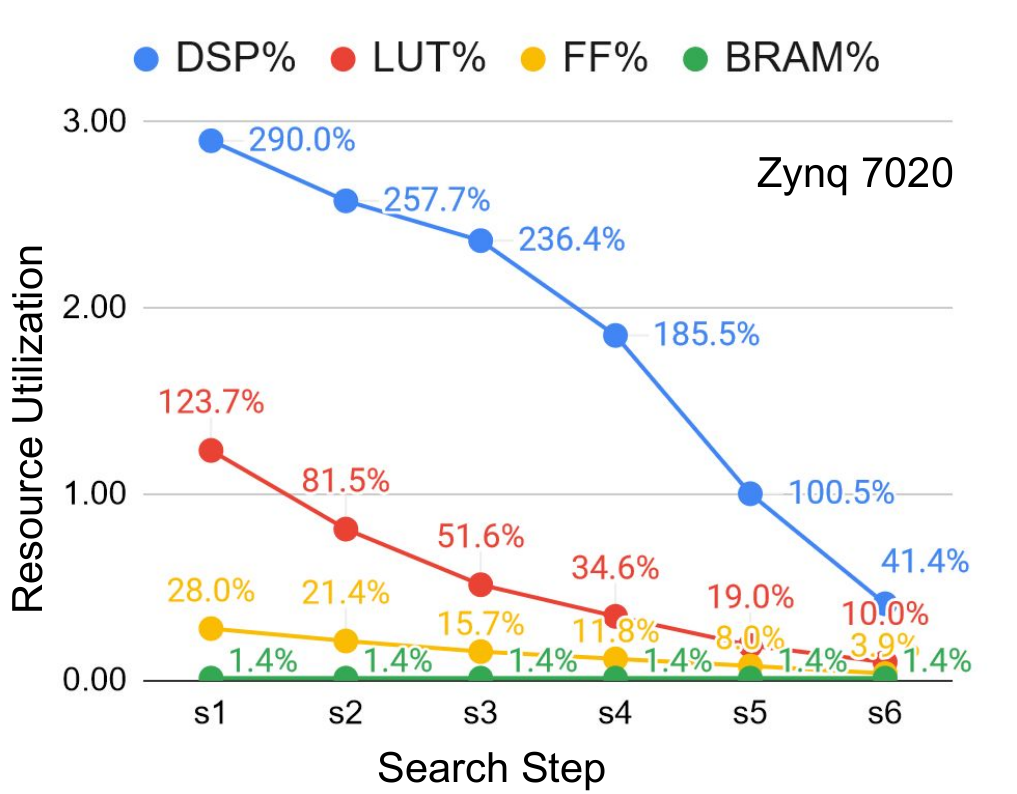}}
   \hspace*{\fill}
  \subfloat[]{%
    \includegraphics[width=0.32\linewidth]{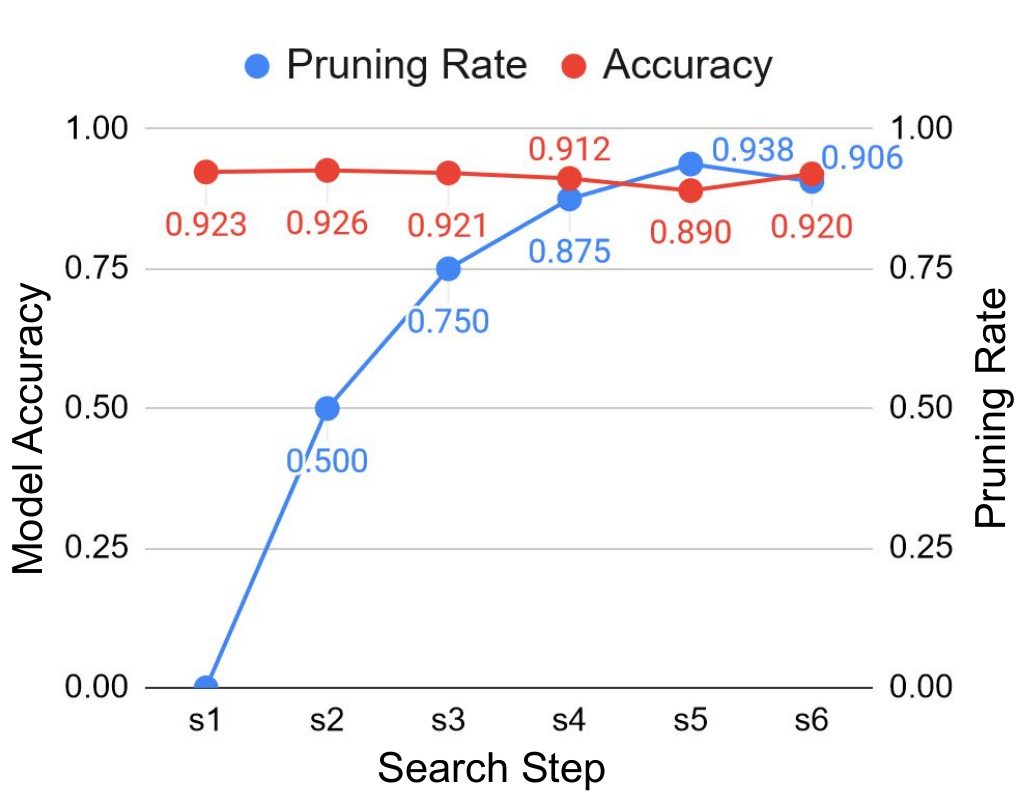}} 
   \\
   \hspace*{\fill}
  \subfloat[]{%
    \includegraphics[width=0.32\linewidth]{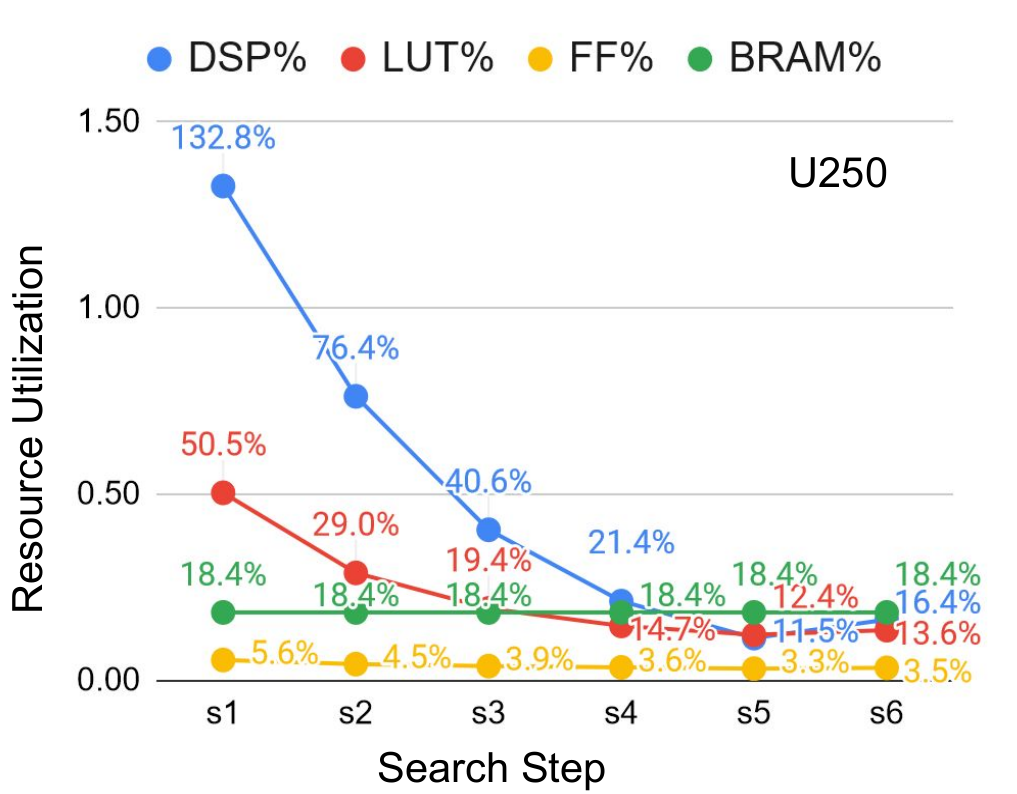}}
   \hspace*{\fill}
  \subfloat[]{%
    \includegraphics[width=0.32\linewidth]{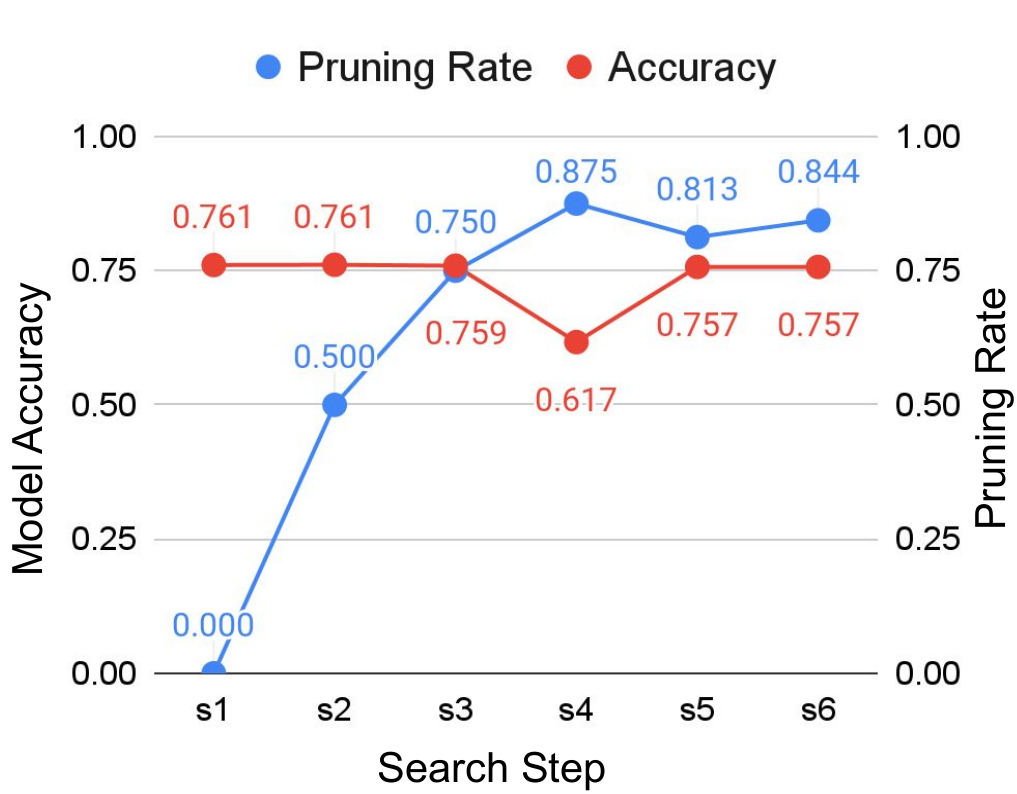}} 
   \hspace*{\fill}
  \subfloat[]{%
    \includegraphics[width=0.32\linewidth]{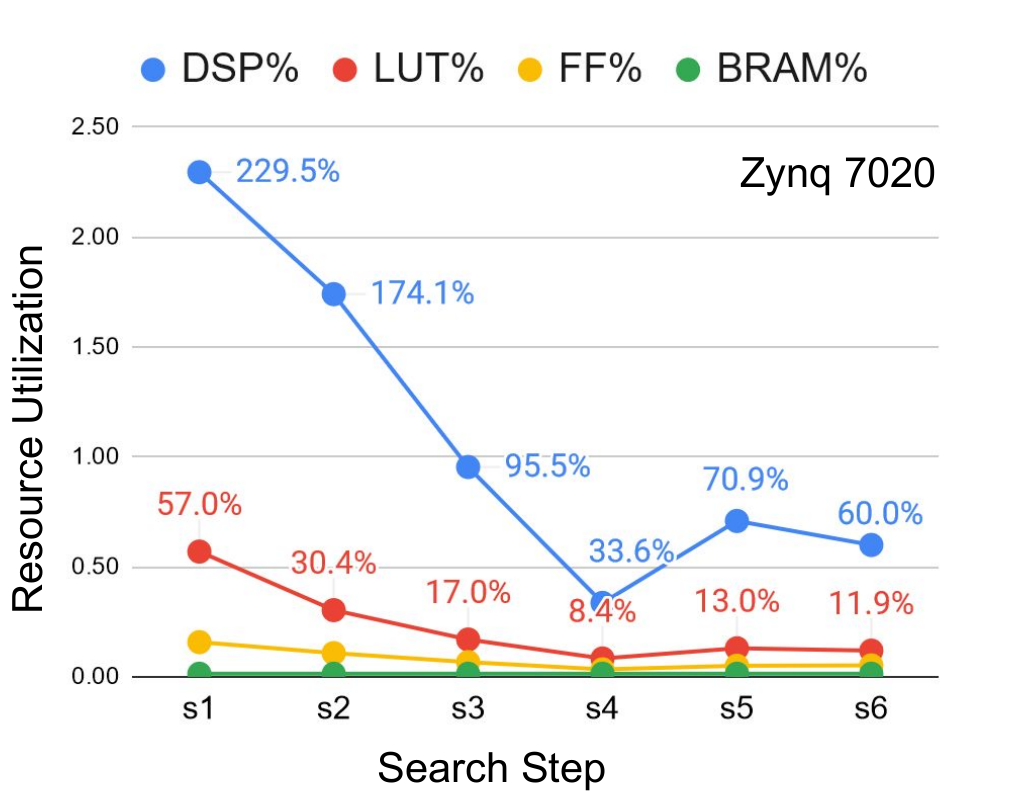}}
   \\
   \hspace*{\fill}
   \subfloat[]{%
    \includegraphics[width=0.32\linewidth]{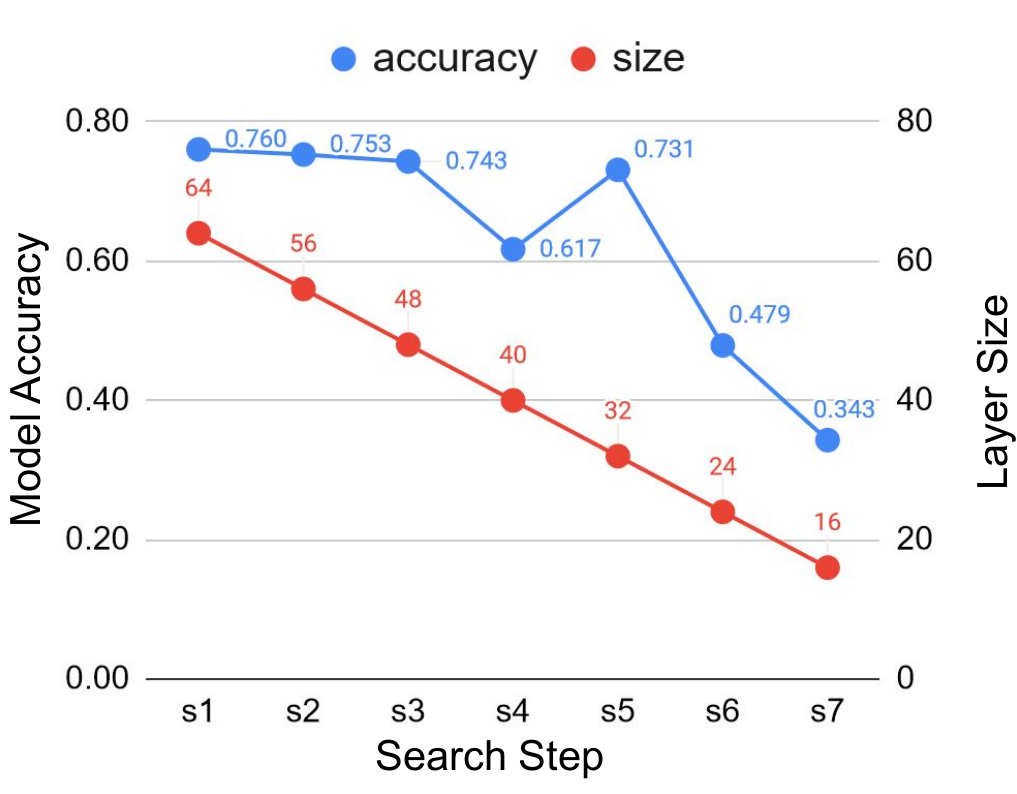}} 
   \hspace*{\fill}
  \subfloat[]{%
    \includegraphics[width=0.32\linewidth]{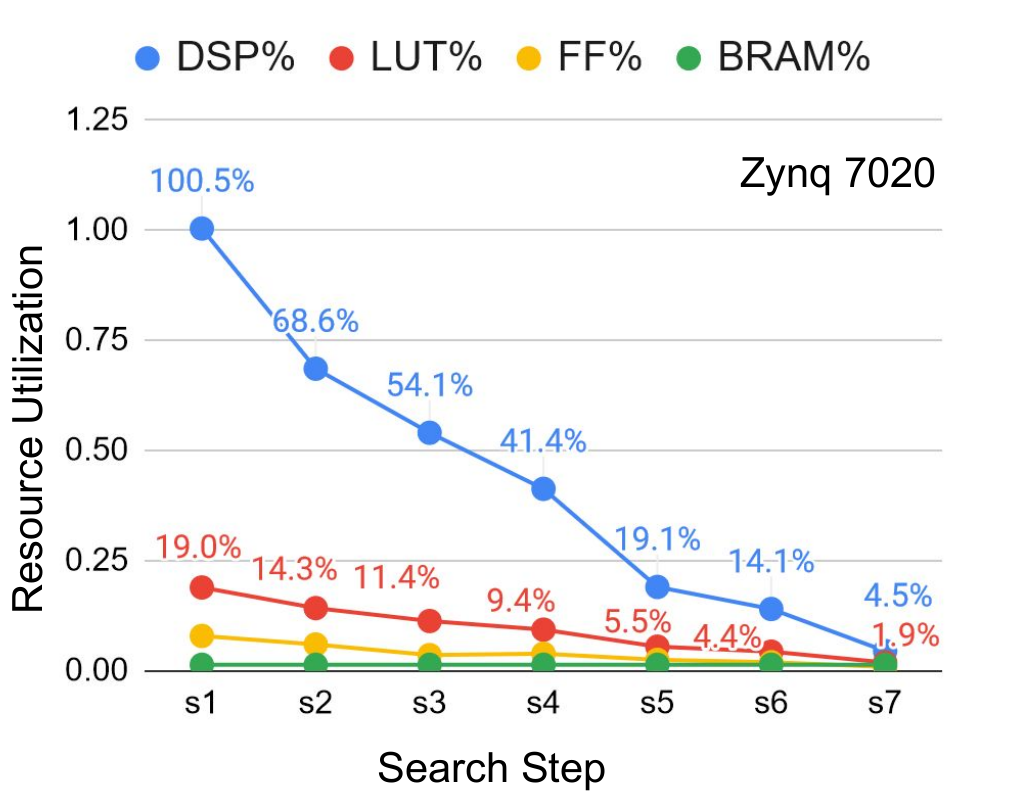}}
    \hspace*{\fill}
  \caption{
  (a) Pruning rates \& accuracy of Jet-DNN.
  (b) Resource utilization of Jet-DNN design candidates after pruning.
  (c) Pruning rates \& accuracy of ResNet9.
  (d) Resource utilization of ResNet9 design candidates after pruning.
  (e) Jet-DNN pruning rates \& accuracy with scaling $\rightarrow$ pruning.
  (f) Resource utilization of Jet-DNN design candidates in (e). 
  (g) Jet-DNN pruning rates \& accuracy with pruning $\rightarrow$ scaling.
  (h) Resource utilization of Jet-DNN design candidates in (g).
  }
  \label{fig:strategy_results} 
\end{figure*}

\begin{figure*} 
   \centering
   \subfloat[]{%
    \includegraphics[width=0.48\linewidth]{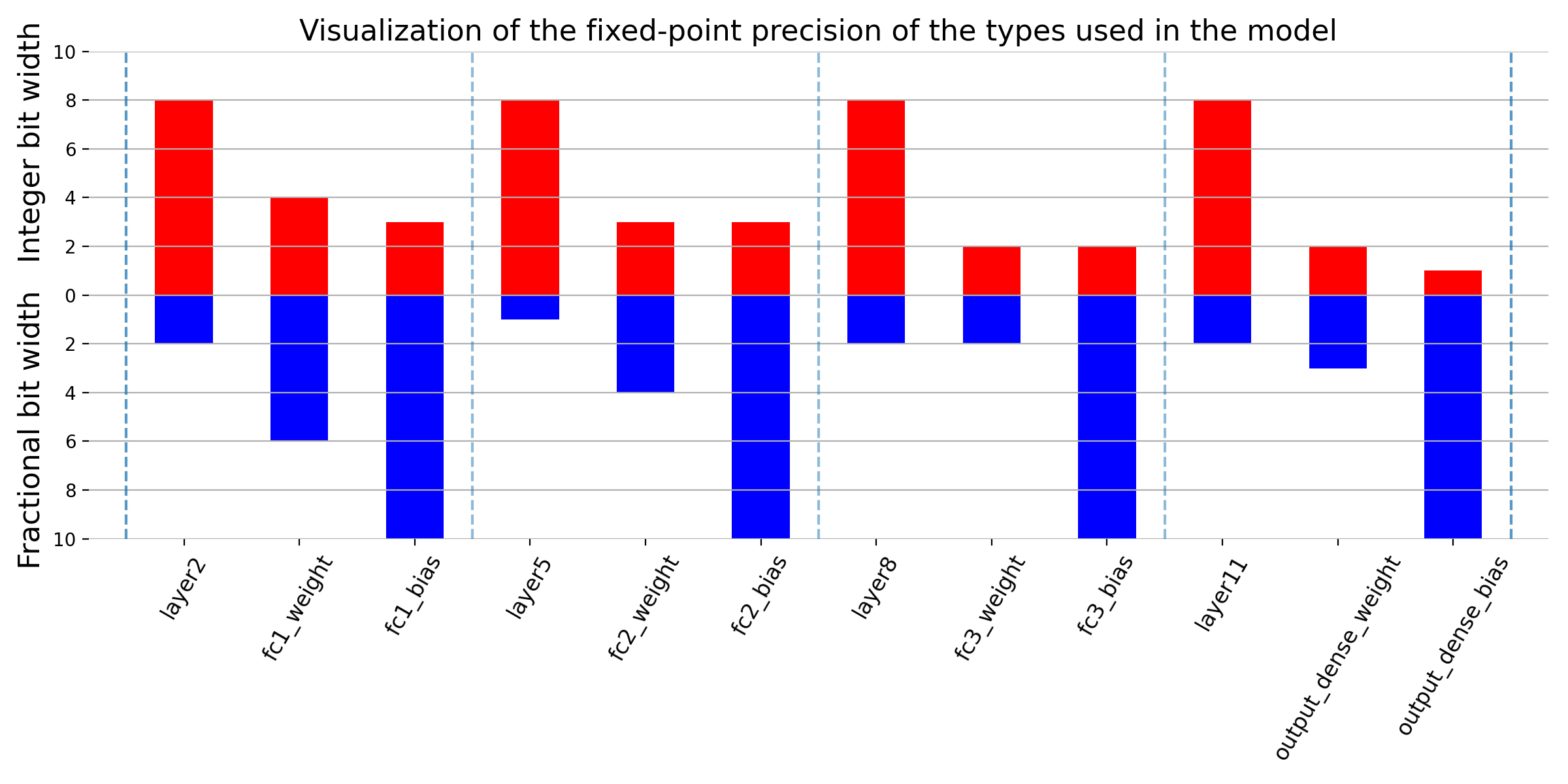}}
  \subfloat[]{%
    \includegraphics[width=0.48\linewidth]{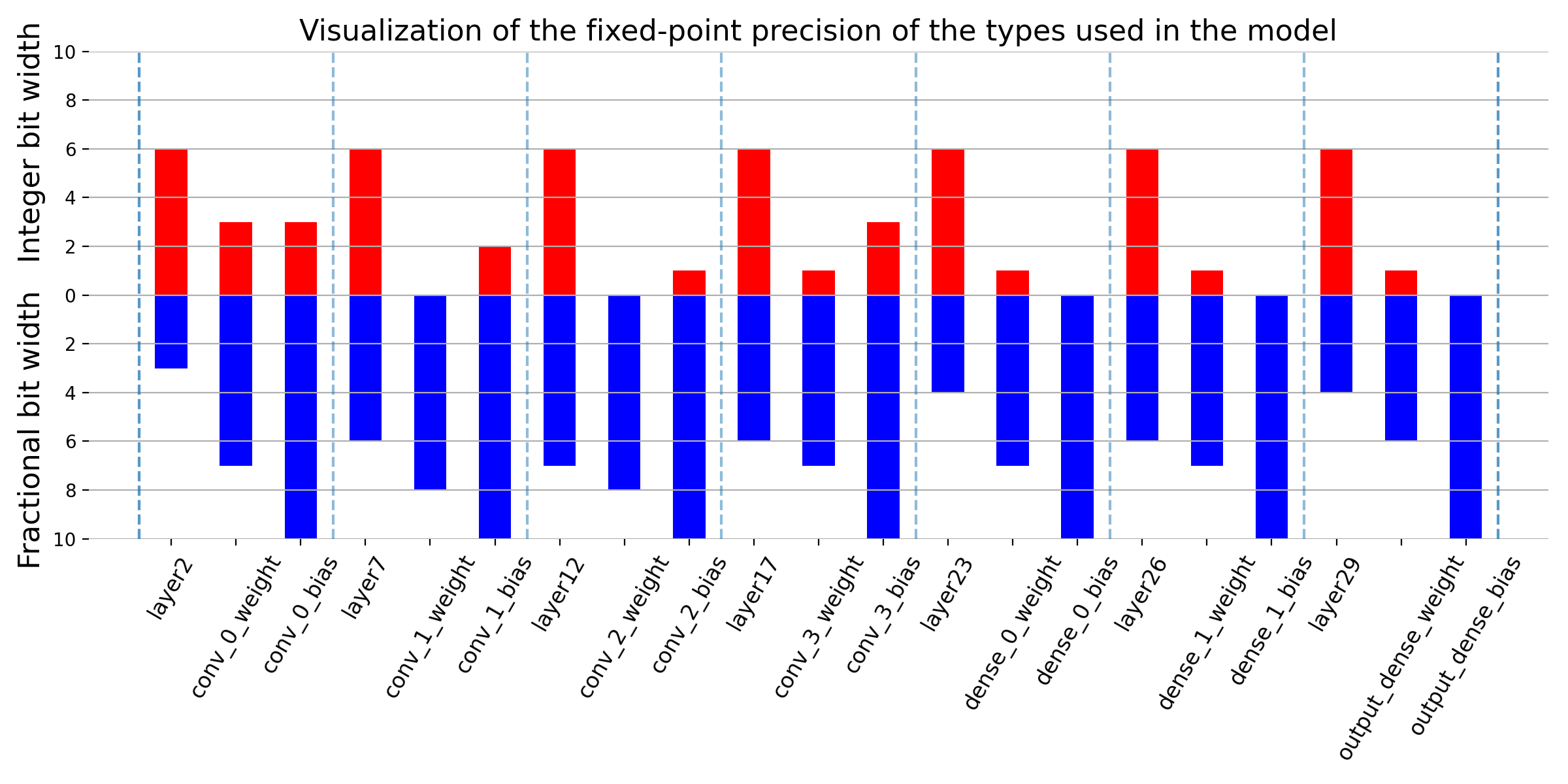}}
 
  \caption{
  (a) Quantized bitwidth of each layer in VGG7. 
  (b) Quantized bitwidth of each layer in JetDNN. 
  }
  \label{fig:quantization_results} 
\end{figure*}

\subsubsection{Pruning strategy.} 

\textcolor{\mycolor}{The design flow for the pruning strategy is illustrated in Fig.~\ref{fig:design_flow}(a) and the search steps and direction are shown in Fig.~\ref{fig:auto_pruning}.
Both the tolerance ($\leq \alpha_{p}$) and threshold ($\beta_{p}$) values are initially set to 2\% in this strategy. }
The effectiveness of the auto-pruning algorithm is demonstrated in Fig.~\ref{fig:strategy_results}. Fig.~\ref{fig:strategy_results} (a) and (c) depict the pruning rate and accuracy for Jet-DNN and ResNet9 in each step, while (b) and (d) show the resources utilization on Zynq 7020 and U250. As the pruning rate increases, hardware resource requirements, particularly DSPs and LUTs, decrease, leading to improved FPGA performance. The design candidate with the highest pruning rate within the allowed tolerance is selected.
\textcolor{\mycolor}{In addition, Fig.~10(c)(d) shows that auto-pruning steadily reduces DSP/LUT utilization on U250 for ResNet-9 while maintaining accuracy within the configured $\alpha_p$, mirroring the Jet-DNN trend and indicating robustness across CNN families.}

\subsubsection{Scaling strategy.} 
To accommodate a large DNN design on an FPGA, we use the SCALING $O$-task that automatically reduces the layer size while tracking the accuracy loss $\alpha_s$. 
The search stops either when the loss exceeds $\alpha_s$.  
If necessary, $\alpha_s$ can be adjusted to achieve further size reduction with minimal impact on accuracy. This work sets $\alpha_s$ to 0.05\%, which allows for model size reduction with negligible accuracy loss.

\begin{table}[t]
\centering

\caption{Results of designs with the quantization strategy using different $\alpha_q$. VGG7 designs are using U250 while JetDNN designs are using Zynq 7020. }
\label{table:results_only_quant}
\scalebox{1.0}{

\begin{threeparttable}
\centering
\begin{tabular}{c  |c|c | c |c |c }
\toprule
 
Model & $\alpha_q$ & DSP & LUT & FF  & Acc \\ 

\midrule

VGG7  Baseline 
&  - 
&  \textbf{4568} (\textbf{37.2\%}) 
&  345k (20.0\%) 
&  65k  (1.9\%) 
& 98.2\%
\\ 
\midrule
VGG7 Quant. 
&  0.01  
&  \textbf{934} (\textbf{7.6\%}) 
&  313k (18.1\%)
&  51k (1.5\%) 
& 97.5\% \\ 

\midrule

VGG7 Quant. 
&  0.05  
&  \textbf{505} (\textbf{4.1\%}) 
&  318k (18.4\%) 
&  66k (1.9\%) 
& 93.9\% \\ 

\midrule
\midrule

JetDNN Baseline~\cite{que2023metaml} 
&  -  
& \textbf{638} (\textbf{290\%}) 
& 66k (124\%) 
& 30k (28\%) 
& 76.1\%
\\ \midrule
JetDNN Quant.
& 0.01  
& \textbf{75} (\textbf{34.1\%}) 
&  57k (107\%) 
&  17k (16\%) 
& 75.5\%
\\ \midrule
JetDNN Quant. 
& 0.05 
&  \textbf{61} (\textbf{27.7\%}) 
&  76k (144\%) 
& 7.2k  (6.8\%)
& 71.6\% \\

\bottomrule
\end{tabular}
\vspace{0.1cm}
\end{threeparttable}
}
\end{table}

\subsubsection{Quantization strategy.} 

This section showcases the evaluation results of the quantization \textbf{(Q)} optimization applied to multiple DNN models within the hardware (HLS) optimization space. \textcolor{\mycolor}{The design flow for the quantization strategy is illustrated in Fig.~\ref{fig:design_flow}(b)}. 

Fig.~\ref{fig:quantization_results} shows the precision of the weights, biases and output of each virtual layer of the VGG7 model after being tuned by the quantization strategy with $\alpha_q$ set to 1\%. 
Table~\ref{table:results_only_quant} shows how the quantization affected key evaluation metrics relating to the performance and resource usage of 2 DNN designs. 
With $\alpha_q$ set to 1\%, the proposed QHS quantization algorithm reduces DSP usage by a factor of 4.9 for the VGG7 model, and 8.5 for the JetDNN model. When $\alpha_q$ is increased to 5\%, the designs are further compacted, but with a larger real accuracy loss. This table highlights how varying $\alpha_q$ levels affect model accuracy, DSP, LUT, and FF usage, illustrating the trade-offs between resource savings and accuracy as well as the effectiveness of the proposed QHS quantization algorithm.

\subsection{Multiple $O$-task Strategies}\label{sec:multi_opt}

With our framework, new strategies can be derived by building and revising a design-flow architecture. For instance, by inserting a scaling $O$-task before the pruning $O$-task in Fig.~\ref{fig:design_flow}(a), a custom combined strategy, \textcolor{\mycolor}{which conducts scaling-then-pruning (scaling $\rightarrow$ pruning),}  can be created with results shown in \figref{fig:strategy_results}(e) and (f). The new optimal pruning rate is 84.4\%, lower than the previous 93.8\% (\figref{fig:strategy_results}(a)), due to reduced redundancy from the preceding scaling task. By switching the order of the $O$-tasks, a different optimization strategy performing pruning-then-scaling (pruning $\rightarrow$ scaling) is achieved, resulting in a 0.7\% accuracy drop after one scaling step, as seen in \figref{fig:strategy_results}(g). Moreover, the three optimization $O$-tasks, pruning, scaling, and quantization, can be integrated into a single automated cross-stage strategy to enhance both performance and hardware efficiency, as illustrated in \figref{fig:design_flow}(c). We discuss the effects of different design-flow architectures with various combinations and orders in Section~\ref{sec:opt_search}, and various tolerable loss in Section~\ref{sec:opt_search_t}.

\begin{figure} 
   \centering
   \hspace*{\fill}
  \subfloat[]{%
\includegraphics[width=0.42\linewidth]{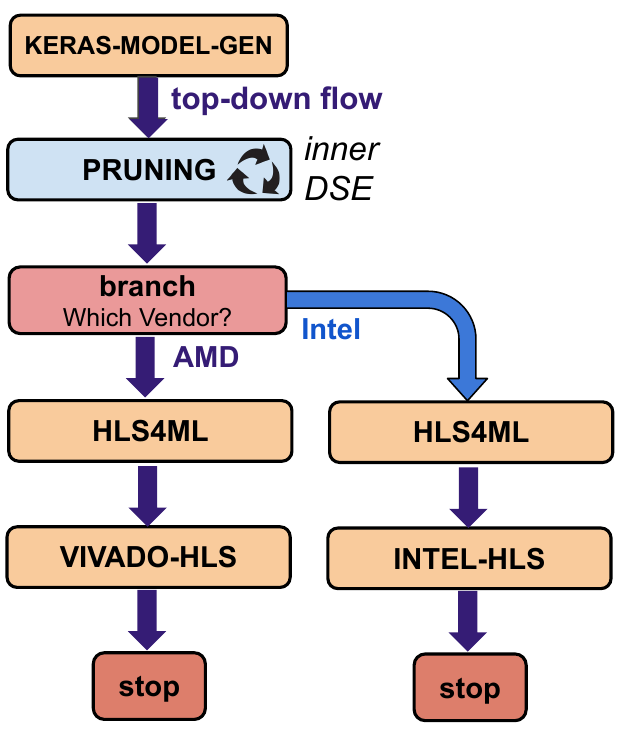}}
   \hspace*{\fill}
   \subfloat[]{%
    \includegraphics[width=0.42\linewidth]{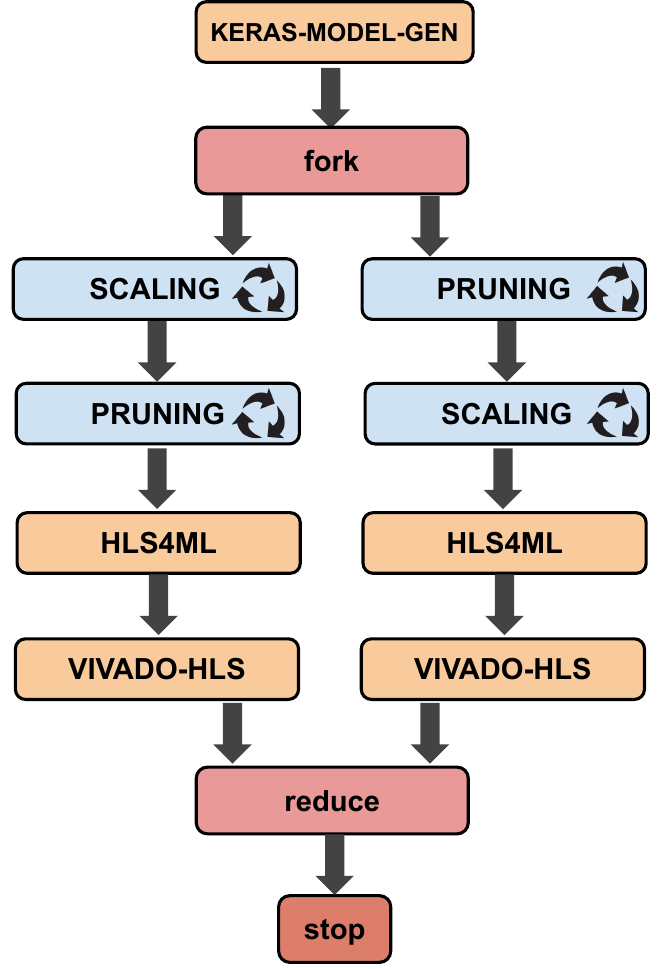}}
 \hspace*{\fill}
  \caption{(a) Pruning optimization targeting different vendors using the BRANCH \textit{K}-task; (b) Combined strategy of scaling and pruning, exploring the order of $O$-tasks. }
  \label{fig:branch_parallel_design_flows} 
\end{figure}

\subsection{Branching Flow}\label{sec:branch}

Fig.~\ref{fig:branch_parallel_design_flows}(a) illustrates a design flow that performs pruning for two alternate targets using the BRANCH $K$-task, for AMD/Xilinx FPGAs and Intel FPGAs. A user-defined selection function is supplied as a parameter to the BRANCH $K$-task which encodes a strategy determining the path forward for the design. 

\begin{figure}
\begin{center}
\includegraphics[width=0.6\linewidth]{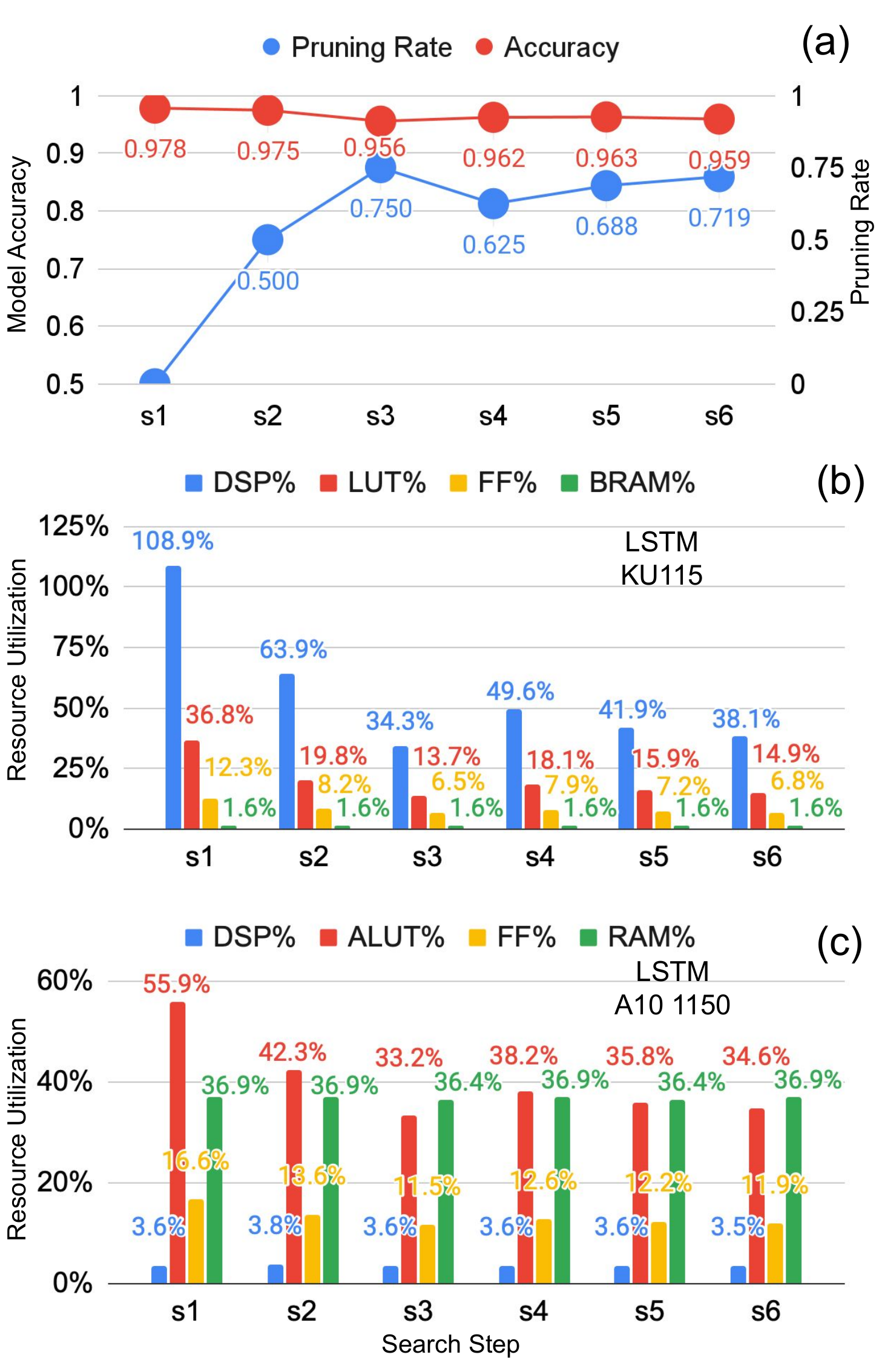}
\end{center}
\caption{LSTM model optimization on two FPGA platforms: (a) pruning rate and model accuracy using the PRUNING $O$-task; (b)~resource utilization on an AMD KU115 FPGA; (c)~resource utilization on an Intel A10 1150 FPGA.}

\label{fig:design_flow_branch_results}

\end{figure}

Fig.~\ref{fig:design_flow_branch_results} presents the results of applying this design flow to an LSTM model on the MNIST dataset. Specifically, Fig.~\ref{fig:design_flow_branch_results}(a) illustrates the pruning rate and accuracy at each step using the PRUNING $O$-task. The tolerance is set to less than $\alpha_{p}$ (2\%) in this design.  Figs~\ref{fig:design_flow_branch_results}(b) and (c) show the resource utilization of the LSTM design after each pruning step respectively on an AMD KU115 FPGA and on an Intel A10 1150 FPGA. The DSP consumption is reduced from 6011 (108\%) to 2101 (38.1\%) on the AMD FPGA after the final pruning rate is optimized to be 71.9\%. Compared to AMD's HLS compiler, which prefers DSP blocks, Intel's HLS compiler tends to favor the use of soft multipliers for implementation. As shown in Fig.~\ref{fig:design_flow_branch_results}(c), most of the computation kernels are implemented using logic resources rather than DSP blocks. \textcolor{\mycolor}{It illustrates that the O-task ports across AMD/Intel with vendor-aware resource trade-offs.}

\textcolor{\mycolor}{While this design flow currently supports two types of FPGAs, it can be extended to include additional paths such as ASIC technologies. We leave the support for ASIC targets as our future work.}
Moreover, this evaluation underscores the flexibility of our approach in utilizing the same software optimization task, specifically PRUNING, across multiple hardware targets.

\begin{figure}
\begin{center}
\includegraphics[width=0.8\linewidth]{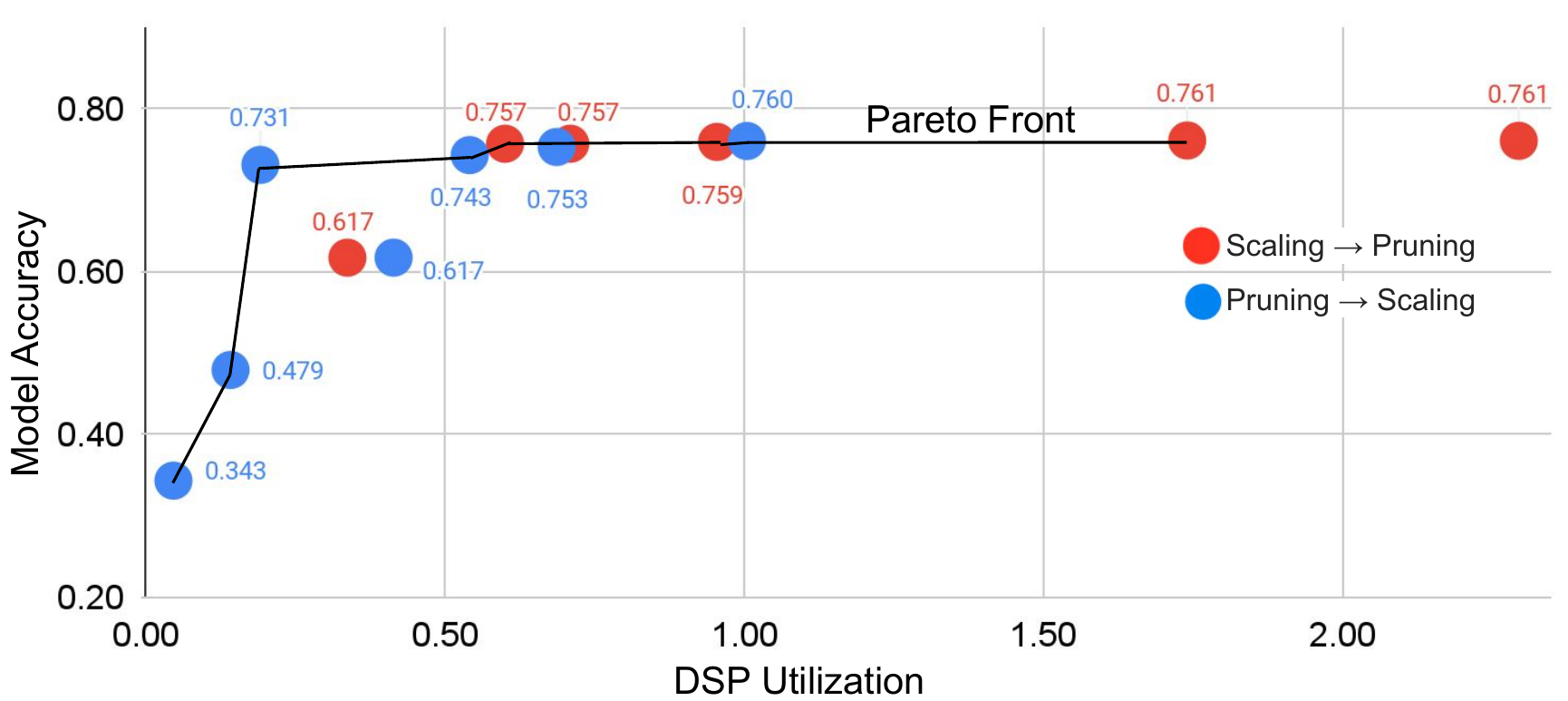}
\end{center}
\caption{Pareto Front meta-model designs post REDUCE $K$-task, color-coded paths.}

\label{fig:parallel_results}

\end{figure}

\subsection{Parallel Flow }
\label{sec:parallel_flow}
Our framework enables the execution of multiple optimization paths in parallel, allowing the selection of the best outcome among them. In Fig.~\ref{fig:branch_parallel_design_flows}(b), we show a design-flow with two parallel paths, where the execution order of two $O$-tasks is changed: scaling $\rightarrow$ pruning and pruning $\rightarrow$ scaling. To support parallel branches, we use the FORK task to connect multiple strategy paths. The results from each path are then evaluated based on predefined criteria, such as accuracy or resource utilization, using the REDUCE task. For this strategy, a Pareto analysis was performed on the designs resulting from both paths, as shown in \figref{fig:parallel_results}. By employing this design-flow, designers can explore various optimization combinations and sequences when the outcomes of these strategies are not clear.

\begin{figure} 
   \centering
   \hspace*{\fill}
  \subfloat[]{%
    \includegraphics[width=0.30\linewidth]{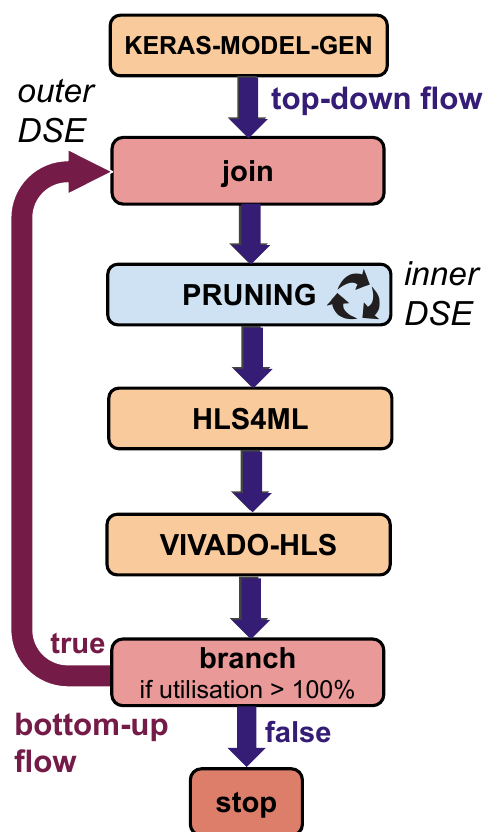}}
   \hspace*{\fill}
   \subfloat[]{%
    \includegraphics[width=0.30\linewidth]{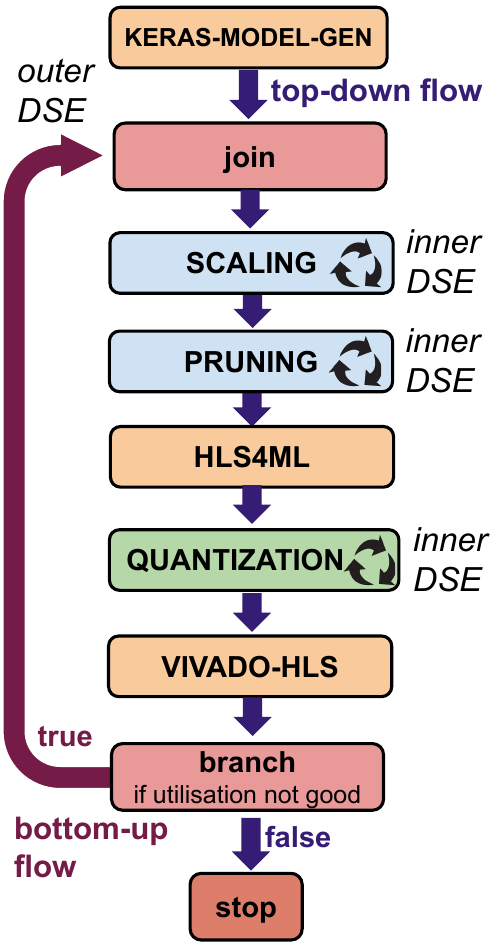}}
  \hspace*{\fill}
   \subfloat[]{%
    \includegraphics[width=0.30\linewidth]{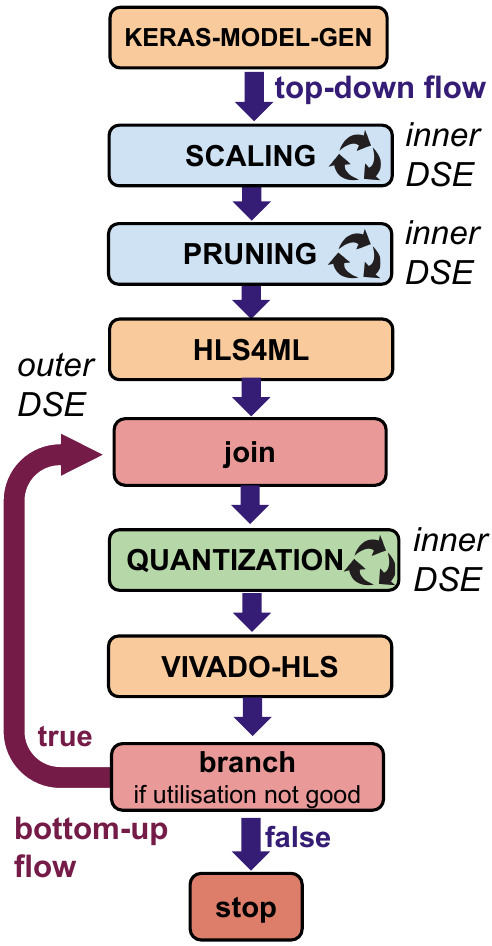}} 
 \hspace*{\fill}
  \caption{(a) Pruning strategy with Branch. (b) Combined strategy with Branch to python (SW) space. (c) The combined strategy with Branch to HLS space. }
  \label{fig:bottom_up_design_flow} 
\end{figure}

\subsection{Bottom-up Flow }
\label{sec:bottom_up}
Fig.~\ref{fig:bottom_up_design_flow} reveal two distinct flows in the optimization strategies: top-down, where information flows from the DNN to the hardware stage, and bottom-up, the reverse. We have automated both flows by customizing the $K$-task BRANCH with a user-defined predicate function to activate or stop the bottom-up flow if the resulting design overmaps. The BRANCH task also supports a user-supplied action function that is triggered when the predicate condition is true. For our current strategies, the action function changes the \textbf{CFG} section of the meta-model, increasing the accuracy tolerance parameters $\alpha_{p}$, $\alpha_{s}$, and $\alpha_{q}$ (if applicable) for the next iteration. Note that our current strategies in Fig.~\ref{fig:bottom_up_design_flow} have two DSE loops: an inner-loop that is codified within each $O$-task and an outer-loop supported by the bottom-up flow. Users can develop more complex strategies by customizing the outer loop, for instance, by updating the bottom-up condition and parameter tuning.

\begin{figure} 
   \centering
   \hspace*{\fill}
  \subfloat[]{%
    \includegraphics[width=0.8\linewidth]{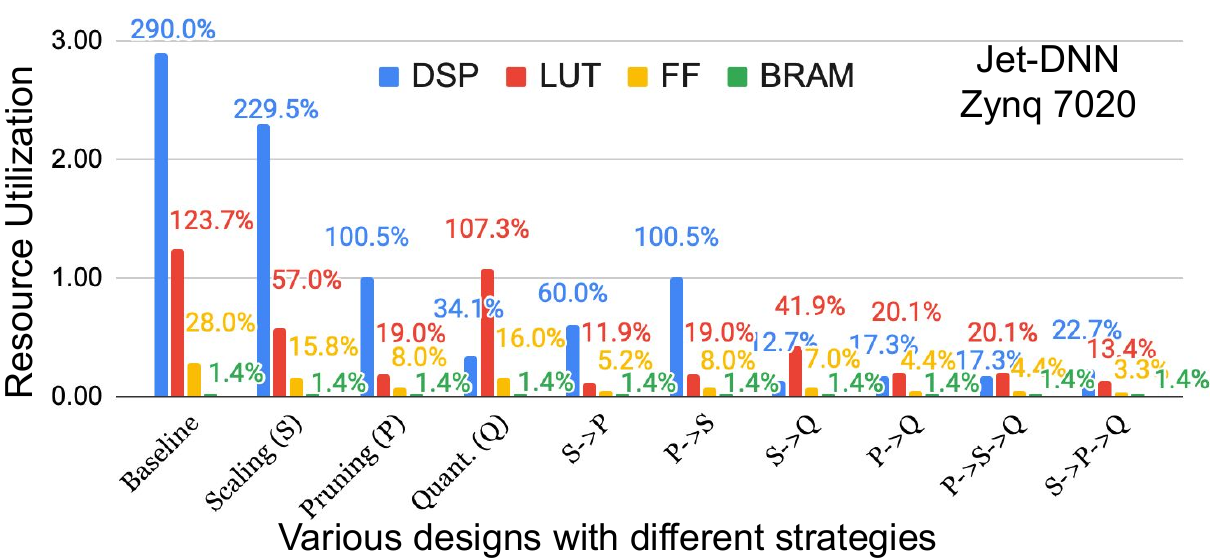}}
   \hspace*{\fill}
   \\
   \hspace*{\fill}
   \subfloat[]{%
    \includegraphics[width=0.8\linewidth]{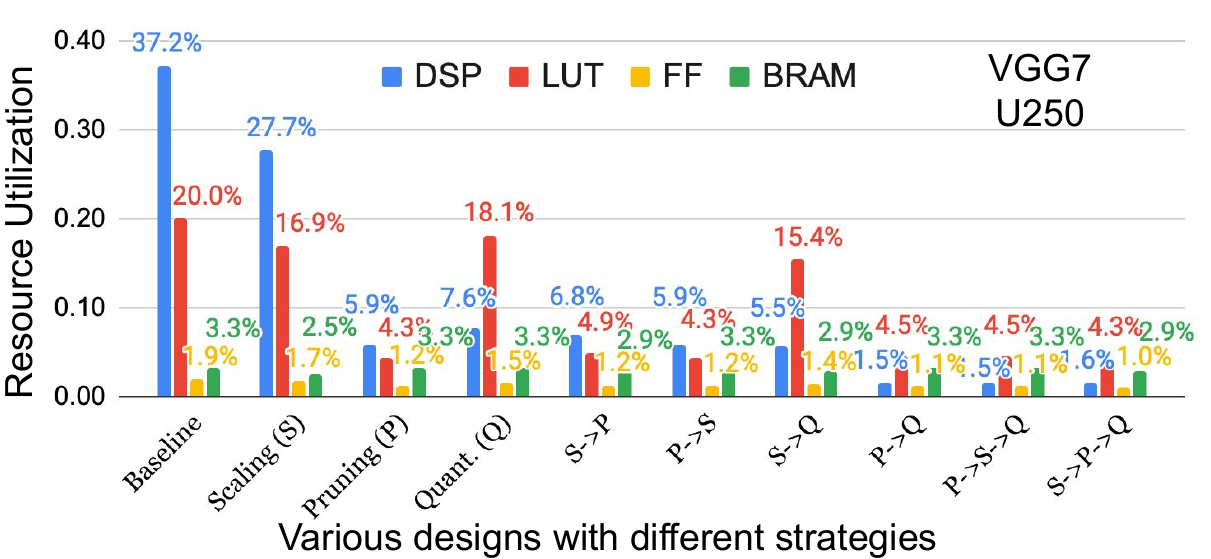}}
  \hspace*{\fill}
  \caption{(a) The hardware resources and latency of the Jet-DNN designs after various strategies. (b) The hardware resources and latency of the VGG7 designs after various strategies.}
  \label{fig:dnn_vgg7_all_resource} 
\end{figure}

\begin{figure}
\begin{center}
\includegraphics[width=0.65\linewidth]{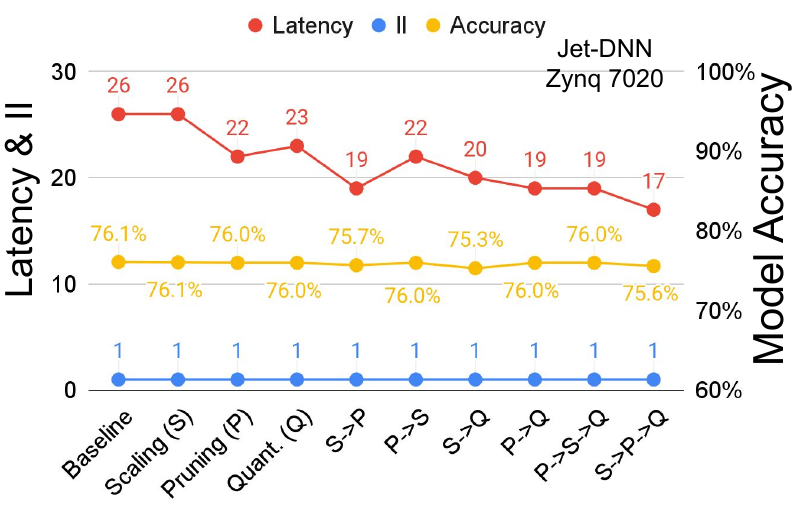}
\end{center}
   \caption{Latency, Initiation Interval (II) and model accuracy of various designs with different strategies}
\label{fig:dnn_all_latency}

\end{figure}

\begin{figure}
\begin{center}
\includegraphics[width=0.80\linewidth]{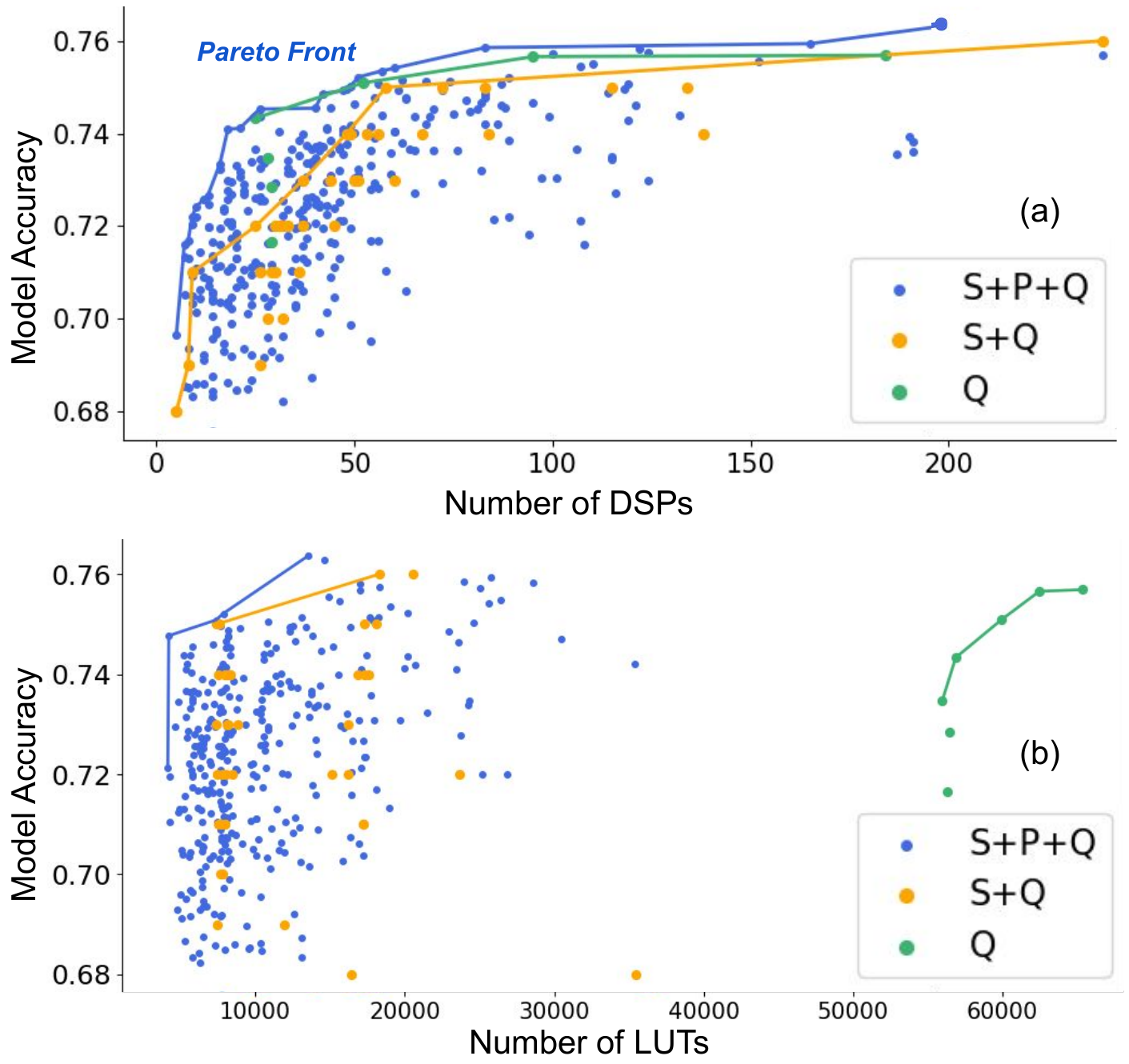}
\end{center}
   \caption{Comparison of Pareto frontiers of \textbf{JetDNN} model accuracy and resource utilization using different optimization strategies.}
\label{fig:strategy_cmp}

\end{figure}

\subsection{Combination and Order}\label{sec:opt_search}

Combining optimization methods increases diversity and potential designs, improving optimal balance between accuracy and efficiency. We systematically evaluate all candidates combined optimization strategies based on our current set of 3 optimization tasks, Scaling (\textbf{S}), Pruning (\textbf{P}), and Quantization (\textbf{Q}), to identify the most effective combination and order. The $\alpha_p, \alpha_s$ and $\alpha_q$ are set to 2\%, 0.05\% and 1\%. 
In particular, Fig.~\ref{fig:dnn_vgg7_all_resource}(a) shows the hardware resource utilization results after each strategy and Fig~\ref{fig:dnn_all_latency} shows the corresponding latency, initiation interval and model accuracy. The final optimized model of Jet-DNN after scaling, pruning and quantization is depicted as ``S$\rightarrow$P$\rightarrow$Q" design, resulting in a reduction of the DSP usage by around 92\% and LUT usage by around 89\% compared with the original design (baseline~\cite{duarte2018fast, que2023metaml}). In addition, the latency is reduced by 35\% while the accuracy loss is trivial, as shown in Fig.~\ref{fig:dnn_all_latency}. 
The same search is performed on the VGG7 network with results shown in~\figref{fig:dnn_vgg7_all_resource}(b). The final design reduces DSP usage by a factor of 23 with the same latency and around 1.1\% accuracy loss compared with the baseline design. 

The search for optimal combinations and sequences with multiple $O$-tasks can be computationally demanding due to the exponential growth of the search space with the number of $O$-tasks. Advanced search methods can replace the current brute-force search. 
Recent studies \cite{kurek2016knowledge, ferretti2022graph, wu2021ironman} 
demonstrate surrogate models can expedite the search. 
Our framework facilitates the deployment of such techniques, enabling users to interface a search algorithm with the hardware design with minimal programming effort.

\subsection{Tolerable Loss Variation}\label{sec:opt_search_t}
This subsection evaluates the performance and efficiency trade-offs when using different values for the maximum tolerable accuracy loss ($\alpha_p, \alpha_s$ and $\alpha_q$) for each optimization task to determine the effectiveness of their combination. 
We compare three strategies: \textbf{Q}, \textbf{S$\rightarrow$Q}, and \textbf{S$\rightarrow$P$\rightarrow$Q}, using Grid Search to create a Pareto frontier by assessing model accuracy against DSP and LUT utilization.
As illustrated in Fig.~\ref{fig:strategy_cmp}, our findings reveal that the \textbf{S$\rightarrow$P$\rightarrow$Q} strategy outperforms both the \textbf{S$\rightarrow$Q} and \textbf{Q} strategies across multiple dimensions, including accuracy, DSP and LUT usage. However, it is important to note that the \textbf{S$\rightarrow$P$\rightarrow$Q} strategy is less efficient in terms of time and design space requirements, demanding 220.5 times more time than the \textbf{Q} strategy. Practical applications may require a balance between performance, hardware resource utilization, and time investment, with the choice of strategy complexity and design space contingent on the specific use case. Our following experiments in this paper build on these findings, focusing on the \textbf{S$\rightarrow$P$\rightarrow$Q} strategy.

\begin{figure}
\begin{center}
\includegraphics[width=0.75\linewidth]{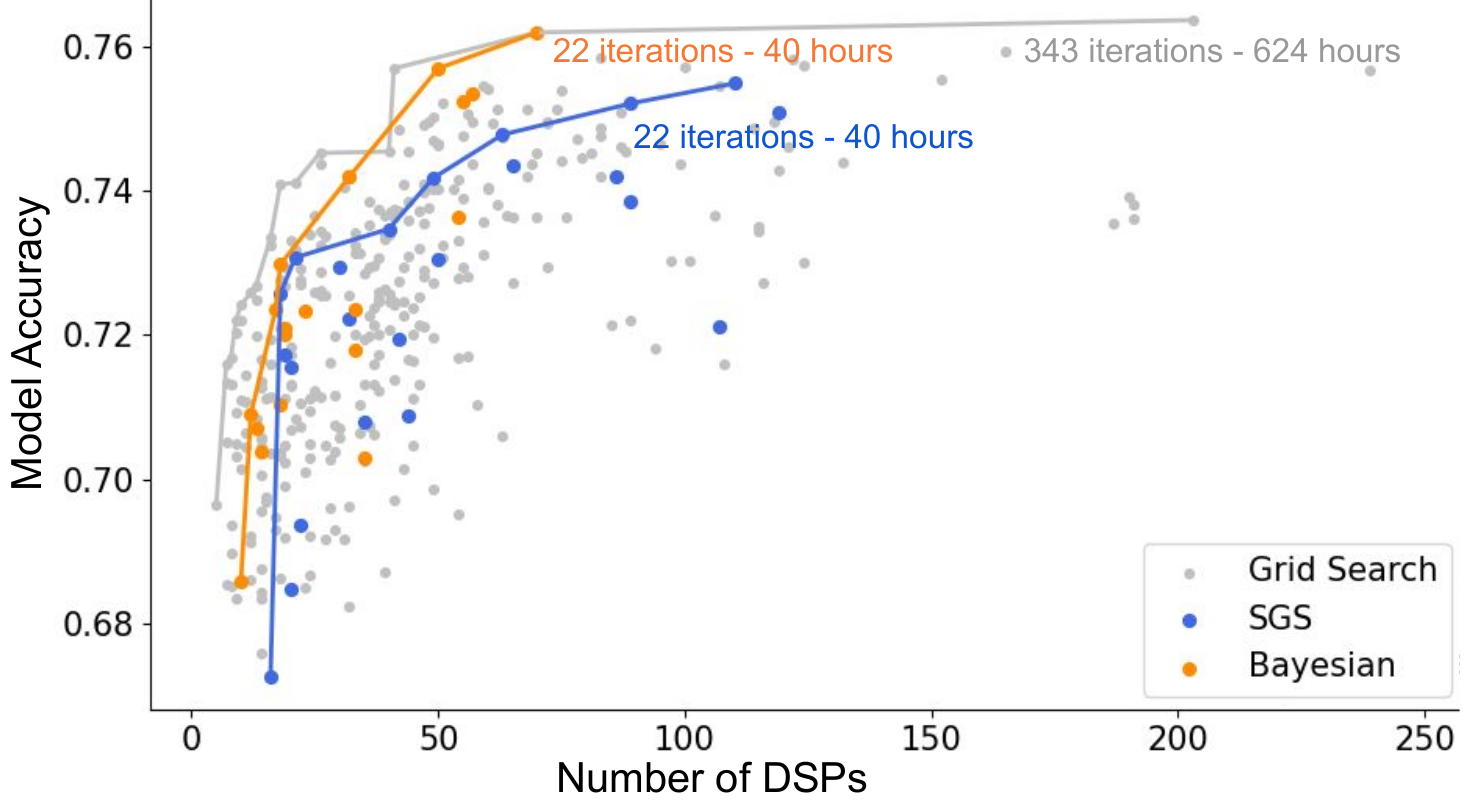}
\end{center}
   \caption{DSP-Accuracy Pareto frontiers for each optimization using different DSE methods for \textbf{JetDNN} models.}
\label{fig:dse_cmp}
\end{figure}

\begin{figure}
\begin{center}
\includegraphics[width=0.75\linewidth]{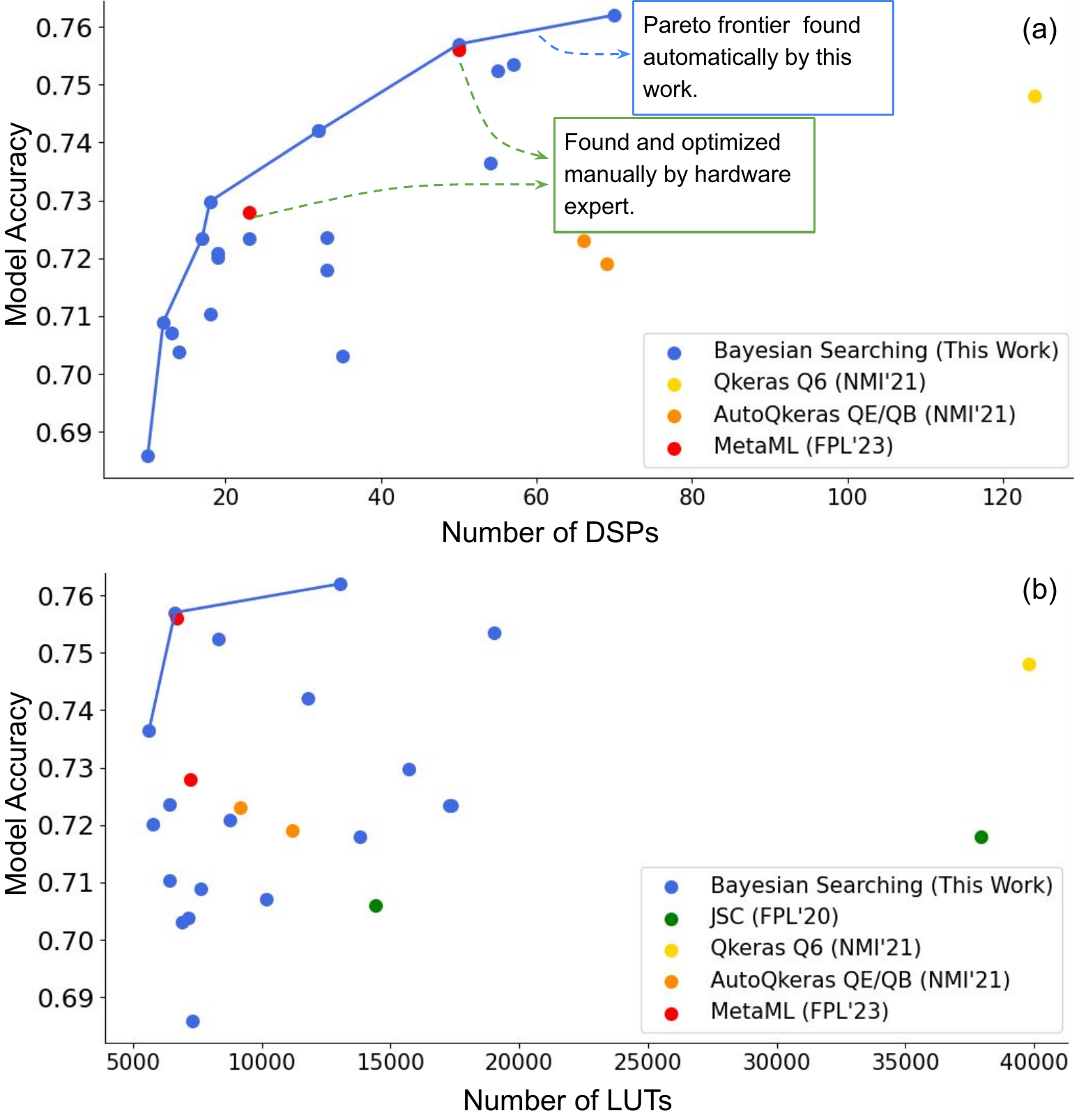}
\end{center}
   \caption{Comparison of resource utilization of the FPGA-based \textbf{JetDNN} networks using our approach and others (LogicNets
JSC~\cite{umuroglu2020logicnets}, Qkeras Q6~\cite{coelho2021automatic}, AutoQkeras QE/QB~\cite{coelho2021automatic} and MetaML~\cite{que2023metaml}) on an AMD/Xilinx VU9P FPGA. The Pareto frontier is highlighted.  }
\label{fig:jetdnn_cmp}
\end{figure}

\begin{table}[t]
\centering

\caption{Performance comparison with the FPGA designs of Jet-DNN network using other approaches on Xilinx FPGAs with a clock frequency of 200MHz. \textcolor{\mycolor}{Please note that this table only reports Jet-DNN to enable an apples-to-apples comparison with prior Jet-DNN FPGA designs.}
}
\label{table:cmp_fpga}
\scalebox{0.85}{
\begin{threeparttable}
\centering
\begin{tabular}{c| c | c |c |c | c |c |c }
\toprule
 
Model 
& $\alpha_s$, $\alpha_p$, $\alpha_q$ (\%) 
& FPGA 
& Acc.(\%) 
& Lat. (ns)
& DSP (\%)
& LUT (\%)
& \newtext{Pow. (W)$^e$}
 \\

\midrule
HLS4ML Jet-DNN~\cite{duarte2018fast} 
& - & KU115 & 75 & 75
& 954 (17.3)
& - 
& - 
 \\
\midrule

FPL'20 LogicNets JSC-M~\cite{umuroglu2020logicnets} 
& - & VU9P & 70.6 & NA & 0 (0) 
& 14,428 (1.2)
& - 
\\
\midrule
FPL'20 LogicNets JSC-L~\cite{umuroglu2020logicnets} 
& - & VU9P & 71.8 & 13$^a$ & 0 (0) 
& 37,931 (3.2)
& - 
\\
\midrule

NMI'21~\cite{coelho2021automatic} Qkeras Q6
& - & VU9P & 74.8 & 55 
& 124 (1.8)
& 39,782 (3.4)
& - 
\\
\midrule

NMI'21~\cite{coelho2021automatic} AutoQkeras QE
& - & VU9P & 72.3 & 55
& 66 (1.0)
& 9,149 (0.8)
& - 
\\
\midrule

NMI'21~\cite{coelho2021automatic} AutoQkeras QB
& - & VU9P & 71.9 & 70
& 69 (1.0)
& 11,193 (0.9)
& - 
\\
\midrule
FPL'23~\cite{que2023metaml} MetaML 
&  - & VU9P &  76.1 & 70 
& 638 (9.3)
& 69,751 (5.9)
& \newtext{2.506}
  \\
\midrule
FPL'23~\cite{que2023metaml} MetaML
&  \{0.05, 2, 1\}  & VU9P &  75.6 &45 
& 50 (0.7)
& 6,698 (0.6)
& \newtext{0.199} 
  \\
\midrule

FPL'23~\cite{que2023metaml} MetaML

& \{0.05, 2, 4\} & VU9P & 72.8 & 40 
& 23 (0.2)
& 7,224 (0.6)
& \newtext{0.169}
 \\

\midrule
\midrule
This work \textbf{(Best Acc.)}
& \{0.5, 0.1, 0.1\} & VU9P & \textbf{76.2} & 55 
& 272 (4.0)
& 14,580 (1.2)
& \newtext{1.096}  
 \\

\midrule
This work \textbf{(Best DSP)} 
& \{0.5, 3, 4\} & VU9P & 69.7 & 40 
& 5 (0.1)
& 9,026 (0.8)
& \newtext{0.118}  
 \\
 
\midrule

This work \textbf{(Best LUT)} 
& \{2, 5, 1\} & VU9P & 72.1 & 45 
& 47 (0.7)
& 4,410 (0.4)
& \newtext{0.204} 
 \\
\midrule

This work$^b$ \textbf{(Acc.-DSP-LUT)}  
& \{0.5, 2, 0.5\} & VU9P & \textbf{76.1} & 50 
& 70 (1.0)
& 13,042 (1.1)
& \newtext{0.386} 
 \\
\midrule

This work$^b$ \textbf{(Acc.-DSP-LUT)}
& \{4, 0.5, 0.5\} & VU9P & \textbf{75.7} & 45 
& 50 (0.7)
& 6,634 (0.6)
& \newtext{0.247} 
 \\

\midrule

This work$^c$ \textbf{(Acc.-LUT)} 
& \{4, 4, 0.5\} & VU9P & \textbf{73.7} & 45 
& 54 (0.8)
& 5,624 (0.5)
& \newtext{0.246} 
 \\

\midrule

This work$^d$ \textbf{(Acc.-DSP)} 
& \{2, 3, 4\} & VU9P & \textbf{70.9} & 40 
& 12 (0.2)
& 7,637 (0.8)
& \newtext{0.130} 
 \\

\bottomrule
\end{tabular}

    \footnotesize
     $^a$ A clock frequency of 384 MHz is used and the final softmax layer is removed. \\
     $^b$ On both Accuracy-DSP and Accuracy-LUT Pareto lines as shown in Fig.~\ref{fig:jetdnn_cmp}. \\
     $^c$ Only on Accuracy-LUT Pareto line. \ \ \ $^d$ Only on Accuracy-DSP Pareto line.  \\ 
     \newtext{$^e$ Dynamic power consumption reported by Vivado out-of-context (OOC) design flow. The static power is about 2.5W for all these designs.} 
    \normalsize
\end{threeparttable}}
\end{table}

\subsection{Design-Space Exploration Strategies}
\label{sec:explore_dse}

Finding the optimal designs with grid search is time-consuming, as discussed in the previous section. This section investigates various DSE algorithms, including grid search, stochastic grid search (SGS), and Bayesian optimization. Fig.~\ref{fig:dse_cmp} presents the results. 
Each colored dot represents a design at a specific iteration, with different colors indicating different algorithms. The solid lines represent the Pareto Frontier of each algorithm over iterations, with each colored line corresponding to 22 iterations taking 40 hours. The uppermost grey line indicates a total of 343 iterations using extensive grid search over 624 hours, representing a baseline for comparison. 
The Bayesian optimization achieves similar results with just 22 iterations, significantly reducing processing time by a factor of 15.6 compared to Grid Search. Compared to SGS, Bayesian optimization's efficient parameter search yields multiple points near the Pareto frontier, indicating its effectiveness in finding optimal designs and approaching global optima. This effective parameter search by employing past results demonstrates the robustness of the Bayesian optimization approach, enabling users to optimize models while minimizing time and effort.

\subsection{Discussion}\label{sec:discussion}

\subsubsection{\textbf{Comparative Evaluation.}} Our evaluation results indicate that our combined $O$-task optimization strategy typically outperforms  single $O$-task techniques. Furthermore, the order in which these optimization techniques are applied plays a crucial role, as different orders produce varying final results, as depicted in~\figref{fig:dnn_vgg7_all_resource}.
 
To highlight the advantages of our framework, we compare our results to those from other approaches targeting low-latency, low-resource, fully unfolded FPGA implementation of the JetDNN network, including LogicNets~\cite{umuroglu2020logicnets} JSC-M and JSC-L, QKeras-based Q6~\cite{coelho2021automatic}, AutoQKeras-based QE and QB~\cite{coelho2021automatic}, 
and MetaML~\cite{que2023metaml} in Fig.~\ref{fig:jetdnn_cmp} and Table~\ref{table:cmp_fpga}.
All designs use the same architecture, except for JSC-L, which employs a larger architecture. 

Compared to original Jet-DNN~\cite{duarte2018fast}, our design achieves higher accuracy (up to 76.2\%) while using less hardware resources. When compared with LogicNets JSC-M and JSC-L~\cite{umuroglu2020logicnets}, which achieve accuracies of 70.6\% and 71.8\% respectively, our design demonstrates up to 5.6\% higher accuracy while also offering resource efficiency. 
Against the AutoQkeras Q6 and QE designs~\cite{coelho2021automatic}, which yield accuracies of 74.8\% and 72.3\% respectively, our framework attains higher accuracy and lower latency while providing more granular trade-offs between resource utilization and latency. 
Compared to MetaML~\cite{que2023metaml} with manual optimization, our work achieves comparable or better accuracy across various configurations with automatic optimization. 
Overall, the versatility of our design, which balances accuracy, latency, and resource utilization across different objectives, highlights its superiority and adaptability for diverse FPGA-based DNN applications.

This effective parameter search demonstrates the robustness of the Bayesian Optimization approach, enabling users to optimize models while minimizing development time. Besides, 
this work demonstrates flexibility and effectiveness by achieving competitive performance across multiple configurations, highlighting its adaptability to different optimization priorities such as Accuracy, DSP and LUT. 
It is worth noting that our results are identified automatically while the other approaches involve manual optimization by DNN and hardware experts.

\subsubsection{\textbf{GNN: Scaling to graph-based models}}

\newtext{Table~\ref{table:cmp_fpga_gnn} shows a case study on the GNN-based jet classifier of Odagiu et al.~\cite{odagiu2024ultrafast} with 8, 16, and 32 constituents on a Xilinx VU13P, where we apply the same PRUNING $O$-task with $\alpha_p=\beta_p=2\%$. MetaML-Pro automatically finds configurations that prune 53.1–65.7\% of the parameters, reducing DSP usage by up to 2.5 times and LUT usage by up to about 2 times, while improving latency from 160–205 ns to 145 ns and the initiation interval from 15 to 10 ns, with accuracy degradations within roughly 1–2 percentage points relative to the baselines. These results extend our evaluation to a larger, irregular graph-based model family.}

\textcolor{\mycolor}{
\subsubsection{\textbf{Additional Exploration Methods.}} Beyond grid and Bayesian search, recent works explore learning-based and reinforcement learning (RL) strategies for accelerator DSE, as well as surrogate/ predictor-driven exploration. Examples include GNN- or RL-assisted HLS DSE and holistic AutoML-style co-design frameworks that learn fast proxies for hardware metrics to guide the search~\cite{wu2021ironman, zhang2020dnnexplorer,reagen2017case,mehrabi2020bayesian,kurek2016knowledge}. 
In our setting, these methods target the same goal as BO-efficiently proposing promising design points under expensive evaluations, but differ in how they trade off sample efficiency, parallelism, and hyperparameter overhead. Our framework is compatible with these alternatives: the outer DSE loop interfaces our cross-stage O-tasks and K-task controls through the meta-model (CFG/LOG/MODEL-SPACE), so an evolutionary or RL agent can be substituted for BO without changing inner optimizations or the flow architecture. 
}

\begin{table}
\centering

\caption{\newtext{
Performance comparison with FPGA designs of GNNs with 8/16/32 constituents on Xilinx VU13P FPGAs with a clock frequency of 200MHz. The pruning strategy is applied with both $\alpha_{p}$ and $\beta_{p}$ set to 2\%. 
}}

\label{table:cmp_fpga_gnn}
\scalebox{0.85}{
\begin{threeparttable}
\centering
\begin{tabular}{c | c| c |c |c | c | c |c |c }
\toprule
 
\newtext{Model} 
& \newtext{FPGA        }     
& \newtext{Pru. (\%)}
& \newtext{Acc.(\%) }
& \newtext{Lat. (ns)}
& \newtext{II (ns)}
& \newtext{DSP (\%)}
& \newtext{LUT (\%)}
& \newtext{Pow. (W)$^a$}
 \\

\midrule
GNN-8C~\cite{odagiu2024ultrafast} 
& VU13P 
& 0
& 64.9 
& 160
& 15
& 2,191 (17.8)
& 472k (27.3)
& - 
 \\
\midrule

GNN-16C~\cite{odagiu2024ultrafast} 
& VU13P 
& 0
& 70.8 
& 180
& 15
& 5,362 (43.6) 
& 1,387k (80.3)
& - 
\\
\midrule

GNN-32C~\cite{odagiu2024ultrafast} 
& VU13P 
& 50
& 75.8 
& 205 
& 15
& 2,120 (17.3) 
& 1,162k (67.3)
& - 
\\

\midrule
\midrule
This work, GNN-8C
& VU13P 
& 65.7
& 63.2 
& 145 
& 10
& 1,005 (8.2) 
& 240k (13.9)
& 5.123 
\\

\midrule
This work, GNN-16C
& VU13P 
& 53.1
& 68.9 
& 145 
& 10
& 5,013 (40.8) 
& 896k (51.9)
& 10.076 
\\

\midrule
This work, GNN-32C
& VU13P 
& 62.5
& 74.6 
& 145 
& 10 
& 845 (6.9) 
& 1,017k (58.8)
& 9.177 
\\

\bottomrule
\end{tabular}

    \footnotesize
     \newtext{$^a$ On-chip power consumption estimated using Xilinx Vivado 2020.1 \textit{report\_power} in out-of-context (OOC) mode at 200 MHz, core logic only.
     } 
    \normalsize
\end{threeparttable}}
\end{table}

\textcolor{\mycolor}{
\subsubsection{\textbf{Extensibility of the Framework.}}\label{sec:extensibility}
While this paper instantiates three SW-level O-tasks (SCALING, PRUNING) and an HLS-level mixed-precision QUANTIZATION (QHS), the pipe-task API and cross-stage controller are agnostic to the specific algorithm. The other optimization, such as hardware-aware NAS, knowledge distillation, structured sparsity and RTL level optimizations can be added as drop-in O-tasks/K-tasks that participate in the same FORK/BRANCH/REDUCE and BO-driven outer loop without changing the framework. In addition to new O/K-tasks, the outer DSE module is swappable: evolutionary or RL-based search can be plugged in as the controller that proposes $(\tau,\Pi)$, orchestrating the same FORK/BRANCH/REDUCE flow and inner searches without changing the pipeline code. 
}

\section{Conclusion}
\label{sec:conclusion}

This paper presents a novel co-optimization framework for FPGA-based DNN accelerators, which comprises building blocks that facilitate rapid development of customized cross-stage design flows, automating the entire design iteration process. The results demonstrate that our approach significantly reduces DSP resource usage by up to 92\% and LUT usage by up to 89\%, while maintaining accuracy and without requiring human effort or domain expertise. In addition, our results reveal that Bayesian optimization in DSE significantly streamlines the process, achieving results comparable to grid search but with a 15.6-fold reduction in processing time. Future work will extend this approach to include RTL space \textcolor{\mycolor}{and ASIC platforms}, support a broader range of DNN architectures, including Variational Autoencoder (VAE)~\cite{que2024low} and large language models (LLMs)~\cite{que2025asicon}, and incorporate more optimization strategies like balancing initiation interval~\cite{que2021accelerating, rognlien2022hardware} and distributed arithmetic (DA)~\cite{sun2025da4ml} for further improvements in hardware efficiency and model performance.

\vspace{0.2cm}
\noindent \textbf{Acknowledgement:} The support of the United Kingdom EPSRC (grant number  
EP/V028251/1,
EP/N031768/1, 
EP/S030069/1, 
and EP/X036006/1), UKRI256,
KIAT, Intel, and AMD is gratefully acknowledged. We thank Markus Rognlien, Filip Wojcicki and Shuo Liu and Anyan Zhao for their help.

\bibliographystyle{ACM-Reference-Format}
\bibliography{main-bib}

\end{document}